\documentclass[aps,amssymb,amsmath,prd,twocolumn,
showpacs,preprintnumbers,superscriptaddress,nofootinbib]{revtex4-2}
\usepackage{graphicx,color}
\usepackage{bm}

\usepackage{comment}

\newcommand{\chiEFT}{$\chi$EFT }

\begin{document}

\preprint{APS/123-QED}

\title{Universal Relations for Neutron Star F-Mode and G-Mode Oscillations}

\author{Tianqi Zhao}
\email{zhaot@ohio.edu}
\affiliation{Department of Physics and Astronomy, Ohio University,
Athens, OH~45701, USA}
\affiliation{Dept. of Physics \& Astronomy, Stony Brook University, Stony Brook, NY 11794-3800}

\author{James M. Lattimer}
\affiliation{Dept. of Physics \& Astronomy, Stony Brook University, Stony Brook, NY 11794-3800}
\email{james.lattimer@stonybrook.edu}

\date{\today}

\begin{abstract}
Among the various oscillation modes of neutron stars, f- and g- modes are the most likely to be ultimately observed in binary neutron star mergers, due to the relatively large coupling and shared frequencies with tidal excitations.
The f-mode is known to correlate in normal neutron stars with their tidal deformability, and therefore with their moment of inertia and quadrupole moment.   Using a piecewise polytropic parameterization scheme to model the uncertain hadronic high-density EOS and a constant sound-speed scheme to model pure quark matter, we refine this correlation and show that these universal relations also apply to both self-bound (quark) stars and hybrid stars containing phase transitions.  We identify a novel 1-node branch of the f-mode that occurs in low-mass hybrid stars in a narrow mass range just beyond the critical mass necessary for a phase transition to appear.  This 1-node branch shows the largest, but still small, deviations from the universal correlation we have found.  It is characterized by a non-monotonic relation between neutron star mass and f-mode frequency, in contrast to the behavior otherwise observed in normal, quark and hybrid stars. 

The g-mode frequency only exists in matter with a non-barotropic equation of state involving temperature, chemical potential or composition (such as being out of beta equilibrium), or a phase transition in barotropic matter.  The g-mode therefore could serve as a probe for studying phase transitions in hybrid stars.  In contrast with the f-mode, g-mode frequencies do not correlate well with tidal deformability, but depend strongly on properties of the transition (the density and the magnitude of the discontinuity) at the transition.  They also weakly depend on the equation of state on either side of the transition.  Imposing causality and maximum mass constraints, the g-mode frequency in hybrid stars is found to have an upper bound of about 1.25 kHz. However, if the sound speed $c_s$ in the inner core at densities above the phase transition density is restricted to $c_s^2\le c^2/3$, the g-mode frequencies can only reach about 0.8 kHz, which are significantly lower than f-mode frequencies, 1.3-2.8 kHz. Also, g-mode gravitational wave damping times are usually extremely long, $>10^4$ s ($10^2$ s) in the inner core with $c_s^2\le c^2/3$ ($c^2$), in comparison with the f-mode damping time, 0.1-1 s.

\end{abstract}

\maketitle


\section{Introduction}

Isolated neutron stars (NSs) are expected to oscillate in many modes, corresponding to different restoring forces. In asteroseismology, fluid oscillations, including g-, f- and p-modes, have been extensively studied in Newtonian gravity \cite{cox2017theory}. In this paper, we focus on f- and g-modes under linearized general relativity formalism \cite{thorne1967non, lindblom1983quadrupole}. The f-mode is the lowest-order, fundamental, non-radial breathing mode, characterized by a zero radial node number $n=0$.  The g-mode is a global oscillation with an arbitrary number of nodes with gravity being the dominant restoring force.  They are a consequence of local buoyancy oscillations, and characterized by small Eulerian pressure variations.  The f- and g-modes we consider have $\ell=2$ so that they may couple with gravitational waves. 

In addition, one could consider also p- and w-modes that have an arbitrary number of nodes.  The restoring force of the p-mode is dominated by pressure variations in matter. The w-mode is the strongly damped gravitational wave (GW) mode dominated by variations of the space-time metric  \cite{kokkotas1992w}. However, p- and w-modes usually cannot be excited during neutron star mergers due to their high frequencies, 5-12 kHz  \cite{kokkotas1999quasi}.  In any case, such high frequencies are effectively unobservable, being well beyond the range of next-generation GW detectors.  We will not consider them further. 

The f-mode is a fundamental mode sitting between the g- and p-modes, with frequency $\nu_f=\omega_f/(2\pi)\sim1.3-2.8$ kHz. It only exists in non-radial modes, $\ell\geq$1. 
The f-mode frequency is known to correlate with the mean NS density $\omega_f\propto\sqrt{G\bar\rho}\propto \sqrt{M/R^3}$~ \cite{andersson1998towards}. However, a more precise, EOS-insensitive, correlation can be found between the dimensionless frequency $\Omega_f=GM\omega_f/c^3$ with other NS properties such as the dimensionless moment of inertia $\bar I=Ic^4/(G^2M^3)$ and the dimensionless tidal deformability $\Lambda$. A semi-universal $\Omega_f-\bar I$ correlation was first proposed by Lau et. al.  \cite{lau2010inferring}, established with a limited number of EOSs. Later, the so-called I-Love-Q relation was discovered \cite{yagi2013love} that provides a rather precise, EOS-insensitive, correlation between $\bar I$, $\Lambda$, and the dimensionless quadrupole moment. Thus, there exists a similar $\Omega_f-\Lambda$ correlation as well  \cite{chan2014multipolar,chirenti2015fundamental,lioutas2021frequency}. In this work, we refine this correlation to encompass any causal EOS constrained by neutron star mass observations and low-density neutron matter studies employing a polytropic parameterization scheme to model matter at supra-nuclear densities.  In particular, we quantify the accuracy and bounds of this correlation and show that it also applies to self-bound (pure quark) and hybrid (hadron-quark) stars.

 There are usually three types of non-zero frequency g-modes corresponding to instabilities when matter moves adiabatically through temperature, chemical composition or density changes.  The local g-mode frequency $\nu_g$ is determined by the Brunt-Vaisala frequency,
\begin{eqnarray}
\nu_g^2&=& g^2\left(\frac{1}{c_e^2}-\frac{1}{c_s^2}\right) e^{\nu-\lambda},
\end{eqnarray}
where $\nu$ and $\lambda$ are the temporal and radial metric functions.  Here, $g=(dp/dr)(\varepsilon+p)^{-1}$ is the local gravity, $p$ and $\varepsilon$ are the pressure and energy density, respectively,  $c_e=\sqrt{dp/d\varepsilon}$ is the equilibrium sound speed and $c_s=\sqrt{\gamma p(\mu n_B)^{-1}}$ is the adiabatic sound speed. $\gamma=(n_B/p)\partial p/\partial n_B$ is the adiabatic index defining how the matter reacts to the adiabatic compression. $\mu=\partial\varepsilon/\partial n_B$ and $n_B$ are the chemical potential and baryon number density, respectively. $g$-mode buoyancy oscillations are stable when $\nu_g^2>0$, while $\nu_g^2<0$ corresponds to a convective region. When matter is marginally buoyant, the g-mode has zero frequency. When the thermal  \cite{reisenegger1992new} or chemical  \cite{wei2020lifting,jaikumar2021g} relaxation time scale is longer than oscillation period, the sound speed in temperature and chemical equilibrium is different from that of the moving, perturbed, matter. For the thermal and chemical g-modes, an arbitrary number of nodes exist, and the principle g-modes with n=1 have the highest frequency. An universal relation between chemical g-mode frequency and lepton fraction was discovered recently \cite{zhao2022quasi} providing key information of symmetry energy at high density. However, these frequencies are relatively small ($\lesssim 0.6$ kHz) in most cases, except for exotic matter involving quarks and hyperons \cite{zhao2022quasi,dommes2016oscillations}.  A g-mode due to a density discontinuity from a phase transition can be understand as a special version of a g-mode due to chemical composition changes, since matter on the low-density side can be treated as having a different composition from that on the high-density side. This situation occurs when matter does not instantaneously change phase upon passing through the phase transition boundary. For this reason, this type of g-mode is also known as an $i$-mode (interface mode) in the literature \cite{mcdermott1988nonradial}. Higher order g-modes do not exist, unless there are multiple density discontinuities. Both the frequency and damping times of an $i$-mode are significantly larger than that of other g-modes and are more likely to be observed in gravitation wave observations  \cite{lau2021probing}. When there is more than one phase transition (density discontinuity) in a NS, multiple groups of g-modes could exist  \cite{rodriguez2021hybrid}. We will focus on the particular type of g-mode, which we will call the discontinuous g-mode, with lowest order $n=1$.

Quadruple oscillations ($\ell=2$) of all modes can couple to and lead to emission of GW radiation, and will be characterized by the GW damping timescale $\tau$. It is of interest to estimate the observability of this radiation.  The amplitude of observed oscillations with frequency $\omega$ is  \cite{benhar2004gravitational}
\begin{equation}
h(t) = h_0 e^{-t/\tau} \cos{\omega t},
\end{equation}
where $h_0=h(0)$.  The observed GW energy flux is
\begin{eqnarray}
F(t)&=&\frac{c^3\omega^2h_0^2}{16\pi G}e^{-2t/\tau}\\
&=& 3.17 e^{-2t/\tau} \left(\frac{\nu}{\rm kHz}\right)^2\left(\frac{h_0}{10^{-22}}\right)^2 \textrm{~ergs~cm}^{-2}{\textrm~s}^{-1},
\end{eqnarray}
The total GW energy is
$E=4\pi D^2\int_0^\infty Fdt$, where the source distance is $D$, or \begin{eqnarray}
E&=&\frac{c^3\omega^2h_0^2 \tau D^2}{8G}\\
&=&4.27\times 10^{49} \left(\frac{\nu}{\rm kHz}\right)^2\left(\frac{h_0}{10^{-23}}\right)^2\left(\frac{\tau}{0.1\textrm{~s}}\right)\left(\frac{D}{\textrm{15~Mpc}}\right)^2 \textrm{ergs}.
\end{eqnarray}
The total radiated energy in corresponding oscillations should be larger than this energy, since bulk viscosity also contributes to dissipation. We use this formula as a lower limit for the radiated energy in this oscillation mode in order to be conservative. To scale this relation, we use a typical frequency $\nu=1$ kHz, damping time $\tau=0.1$ s, and distance $D=15$ Mpc (the distance to the Virgo Cluster). The Advanced LIGO noise power spectrum (sensitivity) at that frequency is about $S_n=4\times 10^{-24}$. Thus, to be potentially observable in Advanced LIGO with a signal-to-noise ratio $\sim2.5$ would require $h_0\sim10^{-23}$ and a total radiated GW energy $E\sim4\times10^{49}$ ergs. With a next-generation instrument such as Cosmic Explorer and Einstein Telescope that has about 10 times the sensitivity of Advanced LIGO, the threshold values become $h_0\sim10^{-24}$ and $E\sim4\times10^{47}$ erg.

There are multiple scenarios for pumping energy into NS oscillations, including core-collapse supernovae, NS mergers, close encounters of a NS with a black hole (BH), and NS starquakes  \cite{kokkotas1999quasi}.
The requisite energy can be compared with the results of  hydrodynamic simulations, which offer powerful methods to study oscillations of proto-NSs from core-collapse supernovae and of rapidly rotating supramassive NSs from NS mergers  \cite{rezzolla2016gravitational}. Core-collapse supernovae have long been considered as promising GW sources  \cite{ferrari2003gravitational}. f- and g-mode oscillations can be identified in recent simulations and show the total GW energy in core-collapse supernovae of order $10^{44}-10^{47}$ ergs depending on the mass and rotation rate of the progenitor  \cite{radice2019characterizing}. Thus, only galactic sources with $D<20$ kpc are likely to be observed in Advanced LIGO observations, at a rate of, at best, a few per century \cite{li2011nearby}. Next-generation instruments such as Cosmic Explore improves the distance threshold to $D<200$ kpc, but won't change the observed rate much since additional large galaxies lie outside this distance.  Since a proto-NS is hot and opaque to neutrinos, its mean density is smaller and their f-mode frequencies should be smaller than that of a cold NS.

For comparison, rapidly rotating supramassive NS remnants from NS mergers have masses likely larger  than $M_{\rm max}$, the maximum mass of cold, non-rotating NS, being temporarily supported by rotation acquired from the binary's orbital angular momentum  \cite{margalit2017constraining}. Due to their large expected ellipticities and oscillation amplitudes, a GW energy from $10^{-2}M_\odot c^2$ to $10^{-3}M_\odot c^2$  could be emitted within 5 ms  \cite{oechslin2007gravitational}. In this case, the observable distance range for an advanced LIGO signal-to-noise ratio of 2.5 is estimated to be $D\lesssim 20-45$ Mpc for Advanced LIGO observations  \cite{shibata2005merger,bauswein2012measuring}.  The binary neutron star merger rate has been  estimated to be $320_{-240}^{+490}$ Gpc$^{-3}$ yr$^{-1}$  based on the O1 and O2  LIGO–Virgo observation runs as well as on the first half of the O3 run \cite{abbott2021population}. Thus, the predicted event rate becomes more favorable, ranging from $6\times 10^{-4}$ yr$^{-1}$ to 0.04 yr$^{-1}$.  With next-generation instruments such as Cosmic Explorer \cite{reitze2019cosmic}, the predicted event rate improves further to 0.06 yr$^{-1}$ to 4 yr$^{-1}$, which now becomes reasonable. 

Gravitational radiation observed in the post-merger phase is complicated by spin-oscillation interactions.
Neutron star merger simulations show that the dominant fluid oscillation of a supramassive NS coincides with the $m=2$ f-mode  \cite{stergioulas2011gravitational}, and has a strong correlation with the isolated NS f-mode frequency  \cite{lioutas2021frequency}, especially for equal-mass mergers. In the case of equal-mass mergers, the peak frequency in supramassive NSs is almost equal to that of the non-rotating f-mode frequency of isolated NSs with the same mass as each of the merging components \cite{ng2020gravitational}.

Besides directly observing gravitational waves from post-merger NS oscillations, there might be additional indirect possibilities during the inspiral phase. During the inspiral, quadruple oscillations could be excited by the periodic tidal interaction from a companion, especially when the orbital frequency approaches the oscillation frequency  \cite{ma2020excitation}. Orbital energy  transferred to quadruple oscillation results in dissipation and an extra phase advance in the gravitational waveform, and could have a large effect \cite{wynn1999resonant}. Because the g-mode has a low frequency, tidal interactions could excite g-mode oscillations well before resonances with the f-mode are reached during the last part of the inspiral.

Since the f-mode frequency is much higher than the orbital frequency, resonant excitations of the f-mode are not likely for non-rotating NSs. However, if an inspiralling NS is counter-rotating, f-mode resonant frequencies could be lowered significantly because the relevant frequency is $\omega_f-2\omega_s$, where $\omega_s$ is the spin frequency. The f-mode has a larger coupling with tidal field compared with the g-mode. For a millisecond pulsar, f-mode resonances in this case could cause phase advances up to hundreds of cycles  \cite{wynn1999resonant,steinhoff2021spin}. However, large phase shifts due to a resonance seemed not to have occurred in the case of the recent binary NS merger GW170817, since its waveform is consistent with a low spin prior.  In any case, binary evolution theory favors low spins as well  \cite{abbott2017gw170817}, rendering this scenario as unlikely. 

Interestingly, a crude estimate of f-mode frequencies of neutron stars can be obtained by combining observations with the information described above.  Dynamical tidal effects can be modeled based on an effective-one-body approach, treating tidal deformability and f-mode frequency as key parameters   \cite{hinderer2016effects,steinhoff2016dynamical,schmidt2019frequency}. A lower bound to the f-mode frequency can be estimated from the non-detection of a significant resonance phase shift, while an upper bound can be estimated from the $\Omega_f-\Lambda$ universal relation previously alluded to.
 The resulting 90\% credible interval of f-mode frequency for GW170817 was reported as 1.43 kHz $<\nu_f<2.90$ kHz for the more massive star and 1.48 kHz $<\nu_f<3.18$ kHz \cite{pratten2020gravitational} for the less massive star.  
 
Another source of NS oscillations could be starquakes that lead to the release of the strain energy in the neutron star crust. Starquakes might have a connection with glitches observed in pulsar timing. Glitch models based on superfluid vortex models generally predict a negligible amount of GW radiation  \cite{sidery2010dynamics}. But glitch models that involve starquakes show a transfer of up to $10^{-11}$ of the total NS rotation energy to f-mode oscillations  \cite{keer2015developing}. For a millisecond NS with spin frequency of 0.3 kHz and a dimensionless moment of inertia $\bar I=10$, starquakes could emit up to $10^{40}$ ergs in GW. Therefore, only an event   very close to the Earth ($D\lesssim 2$ kpc) could be observable even in next-generation instruments. The closest known pulsar, RX J1856-3754, is about 0.12 kpc away, which lies within the observable range. Assuming a uniform pulsar distribution implies the existence of about $10^6$ pulsars within the Galaxy.  However, 
only about 3000 pulsars are observed~\cite{sartore2010galactic}, giving an average distance of about 1.3 kpc to the nearest pulsar, and indicating that most remain undetected. 
Based on decades of pulsar monitoring, about 10\% of observed pulsars show glitches \cite{fuentes2017glitch}. If a similar percentage of unobserved pulsars are capable of glitching, at most a few might lie in the observed range of next-generation instruments.  

In this paper, we calculate f-mode and g-mode frequencies and damping times by solving relativistic non-radial oscillation equations  \cite{lindblom1983quadrupole,detweiler1985nonradial} which form an ordinary differential equation (ODE) eigenvalue problem. The f-mode is the mode with zero radial nodes in these calculations, and the g-mode is the mode with a single radial mode having the next-lowest frequency. In such calculations, NSs are usually assumed to be non-rotating, although rotation could slightly increase the f-mode frequency  \cite{kojima1993normal}. Going beyond the slow-rotation limit requires a more complicated solution involving the time evolution with partial differential equations \cite{kruger2020dynamics}, which could be calculated from dynamical non-linear GR simulation \cite{rosofsky2019probing,shashank2021f}. Previous studies have mainly focused on hadronic NS. The few existing calculations involving quark stars and hybrid NSs generally have assumed an MIT bag model with $c_s^2\approx1/3$ for the NS inner core  \cite{yip1999quadrupole,sotani2011signatures,flores2014discriminating}. However, we will study models incorporating higher sound speeds in both bare quark stars and in the inner core of hybrid NSs. We employ a parameterized hadronic EOS omitting temperature and chemical composition dependence, as discussed in Section \ref{sec:EOS}, allowing us to explore a wide variety of hadronic, hybrid and pure quark NSs. 
Since our parameterized EOS doesn't have temperature and chemical composition information, we study NS oscillations in equilibrium (i.e., equal equilibrium and adiabatic sound speeds) for both hadronic and hybrid configurations. A long transition timescale is assumed at quark-hadron interfaces, meaning that particle concentrations are frozen during oscillations at the interface. 

Many previous studies, especially for hybrid NSs and quark stars, have used the Cowling approximation   \cite{mcdermott1983nonradial,ranea2018oscillation,sotani2011signatures,flores2014discriminating,kumar2021non,das2021impacts}, which lacks dissipation due to gravitational waves. The Cowling approximation introduces about a 20-30\% error in the f-mode frequency  \cite{yoshida1997accuracy,sotani2001density,chirenti2015fundamental}, which is significantly less accurate than that of the $\Omega_f-\bar I-\Lambda$ universal relation to be studied  \cite{lau2010inferring,chan2014multipolar,chirenti2015fundamental}. An ~5\% error in the g-mode frequency  \cite{sotani2001density} from the Cowling approximation has also been found in $M\leq 1.2 M_\odot$. Massive NSs with stronger gravity, in principle, result in even larger error in g-mode from the Cowling approximation \cite{zhao2022quasi}. Beside, non-linear numerical simulation could reproduce the f-mode frequency, but the damping times depart from linear perturbation theory \cite{rosofsky2019probing}. Instead, in Sec. \ref{sec:oscillation}, we solve metric perturbation equations together with fluid perturbation equations to study the oscillation mode frequencies and gravitational damping time scales. In Sec. \ref{sec:newtonian} we illustrate this formalism applied in the case of Newtonian geometry to selected analytic equations of state, and compare to general relativistic results with and without Cowling approximation. In Sec. \ref{sec:EOS} we develop a six-parameter hadronic EOS together with a two-parameter extension to describe a hybrid EOS.  Results for the f-mode are presented in Sec. \ref{sec:f-I-Lambda}, where the EOS-insensitive $\Omega_f-\bar I-\Lambda$ relation is constructed.  We determine bounds for these relations and quantify their uncertainties. Other fitting formulae are also described. We describe, for the first time, a special 1-node branch of the f-mode associated with so-called twin stars whose masses are near the transition mass to hybrid stars in Sec. \ref{sec:nodes}.  The discontinuous g-mode, present in stars with a phase transition density discontinuity, is analyzed in Sec. \ref{sec:g-mode}, where an EOS-insensitive correlation with stellar compacntess is developed and quantified.

\section{Oscillation Frequencies and Damping Times \label{sec:oscillation}}

Thorne et. al. first studied NS oscillations coupled with gravitational radiation \cite{thorne1967non}. Oscillations of NS are expected to involve linear variations of matter and metric in various sphercial harmonics. The angular decomposition of variations will contain even and odd parity components. Odd parity variations have a trivial zero mode which corresponds to differential rotation, unless axiel symmetry is broken by rotation which result in r-modes \cite{chandrasekhar1991non}. In this work, we study only even parity perturbation of the Regge-Wheeler metric,
\begin{equation}
\begin{split}
ds^2 = - e^{\nu(r)} (1+r^\ell H_0(r)e^{i\omega t} Y_{\ell m}(\phi,\theta))c^2 dt^2\\
+e^{\lambda(r)} (1-r^\ell H_0(r)e^{i\omega t} Y_{\ell m}(\phi,\theta))dr^2 \\
+ (1-r^\ell K(r)e^{i\omega t}Y_{\ell m}(\phi,\theta))r^2 d\Omega^2\\
-2i\omega r^{\ell+1}H_1(r)e^{i\omega t}Y_{\ell m}(\phi,\theta) dt dr
\end{split}
\end{equation}
where $H_0$, $H_1$, and $K$ are metric perturbation functions. $\omega$ is the complex oscillation frequency; its real component is the oscillation frequency and its imaginary component is the inverse of the damping (growth) time if it's positive (negative).   The metric perturbation functions inside the star must match those outside the star at the stellar surface.

\subsection{Perturbations inside the NS}

Fluid perturbation vectors inside the star can be decomposed in  a basis of spherical harmonics in terms of $Y^\ell_m$, $\partial_\theta Y^\ell_m$ and $\partial_\phi Y^\ell_m$. For non-rotating neutron stars, odd parity fluid perturbations have a trivial solution which corresponds to differential rotation, while fluid perturbations with even parity are described by the Lagrangian displacement vectors
\begin{eqnarray}
\xi^r &=& r^{\ell-1}e^{-{\lambda}/{2}}W Y^\ell_m e^{i\omega t}\nonumber\\
\xi^\theta &=& -r^{\ell-2} V \partial_\theta Y_m^\ell e^{i\omega t}\\
\xi^\phi &=& -\frac{r^{\ell-2}}{ \sin^{2}\theta} V\partial_\phi Y_m^\ell e^{i\omega t},\nonumber
\end{eqnarray}
which define the fluid perturbation amplitudes $W$ and $V$. In case of radial oscillations when $\ell=0$, the angular fluid perturbation $V$ is irrelevant due to the vanishing of $\partial_\theta$ and $\partial_\phi$. Perturbations of a spherical star have four degrees of freedom: three coming from the metric perturbations, which will be reduced by one applying Einstein's equation, $\delta G^{01}=8\pi \delta T^{01}$, and two coming from fluid perturbations.  An additional fluid perturbation amplitude $X$ , related to Lagrangian pressure variations, is defined according to 
\begin{equation}
\Delta p = -r^{\ell}e^{-{\nu}/{2}}X Y_m^\ell e^{i\omega t}\,.
\end{equation}
In order to avoid potential singularity in the eigenvalue problem, Lindblom et. al. pick the four independent variables to be $H_1$, $K$, $W$, $X$ \cite{lindblom1983quadrupole,detweiler1985nonradial} and evaluate the two remaining functions $H_0$ and $V$ according to 
\begin{eqnarray}
H_0&=&\left\{8\pi r^2 e^{-\nu/2}X-\left[(n+1)\textrm{Q}-\omega^2r^2 e^{-(\nu+\lambda)}\right]H_1 \nonumber \right. \\
&+&\left. \left[n-\omega^2 r^2 e^{-\nu}- \textrm{Q}(\textrm{Q}e^{\lambda}-1)  \right] K \right\} (2b+n+\textrm{Q})^{-1}
,\nonumber\\
V&=&\left [ \frac{X}{\varepsilon+p}-\frac{\textrm{Q}}{r^2}e^{(\nu+\lambda)/2}W-e^{\nu/2}\frac{H_0}{2}\right ]\frac{e^{\nu/2}}{\omega^2}\label{eq:V_def_in},
\end{eqnarray}
where $n=(\ell-1)(\ell+2)/2$, $Q=b+4\pi G r^2 p/c^4$ and $b=Gm/(rc^2)$ with $m(r)$ the mass interior to $r$. By expanding Einstein's equation to first-order, homogeneous linear differential equations for $H_1$, $K$, $W$ and $X$ can be found \cite{detweiler1985nonradial},
\begin{eqnarray}
r\frac{dH_1}{dr}&=&-\left[l+1+2 b e^\lambda+4\pi r^2 e^\lambda(p-\varepsilon)\right]H_1\nonumber\\
&&\quad\quad +e^\lambda\left[H_0+K-16\pi(\varepsilon+p)V\right],\nonumber
\\
r\frac{dK}{dr}&=&H_0+(n+1)H_1\nonumber\\
&&+\left[e^\lambda Q-\ell-1\right]K-8\pi(\varepsilon+p)e^{\lambda/2}W,\nonumber
\\
r\frac{dW}{dr}&=&-(\ell+1)\left[W+\ell e^{\lambda/2}V\right]\label{eq:ODE_DL3}\\
&&+r^2 e^{\lambda/2}\left[\frac{X}{ (\varepsilon+p) c_s^2 }e^{-\nu/2}+\frac{H_0}{2}+K\right],\nonumber
\end{eqnarray}
\begin{eqnarray}
r\frac{dX}{dr}&=&-\ell X+\frac{\varepsilon+p}{2}e^{\nu/2}\left\{(3e^\lambda Q-1)K-\frac{4(n+1) e^\lambda Q}{r^2}V\right.\nonumber\\
+(1&-&e^\lambda  Q)H_0+(r^2\omega^2e^{-\nu}+n+1)H_1+\left[2\omega^2 e^{\lambda/2-\nu}\right.\nonumber\\
&-&\left. \left.8\pi(\varepsilon+p)e^{\lambda/2}+r^2\frac{d}{dr} \left(\frac{e^{-\lambda/2}}{r^2} \frac{d\nu}{dr} \right)\right]W \right\}.\nonumber
\end{eqnarray}
where  the adiabatic sound speed $c_s$ of NS matter under oscillation is different from the equilibrium sound speed $c_e$\cite{wei2020lifting,jaikumar2021g}. Here, we only consider zero-temperature EOSs without varying chemical composition, so that $c_s=c_e$. 
The central boundary conditions for the perturbation amplitudes are
\begin{eqnarray}
W(0)&=&1,\nonumber\\
X(0)&=&(\varepsilon_0+p_0)e^{\nu_0/2}\times \nonumber\\
 &&\hspace*{-1cm}\left \{ \left[ \frac{4\pi}{3}(\varepsilon_0+3p_0)-\frac{\omega^2}{\ell} e^{-\nu_0}\right]W(0)
+ \frac{K(0)}{2}\right \}, \label{eq:BC_H1}\\
H_1(0)&=&\frac{\ell K(0)+8\pi(\varepsilon_0+p_0)W(0)}{n+1}.\nonumber
\end{eqnarray}
where $\nu_0=\nu(0)$ and the last boundary condition is achieved by solving the two trial solutions with $K(0)=\pm(\varepsilon_0+p_0)$ and then linearly constructing the correct solution satisfying the outer boundary condition $X(R)=0$ (corresponding to no pressure variations at the surface). Note that $H_0(0)=K(0)$ by construction. For a hybrid NS (see Section \ref{sec:EOS}), we assume no chemical changes at the transition boundary  \cite{finn1987g,tonetto2020discontinuity}. Thus, $H_1$, $K$, $W$, $X$ should all be continuous at the transition, while $H_0$ and $V$ are fixed by Eq. (\ref{eq:V_def_in}).
In this paper, we confine the remainder of our discussion to non-radial oscillations with $\ell=2$, so that $V$ and $W$, which are defined only inside a star, are dimensionless functions.

\begin{figure}
    \centering
    \includegraphics[width=\linewidth]{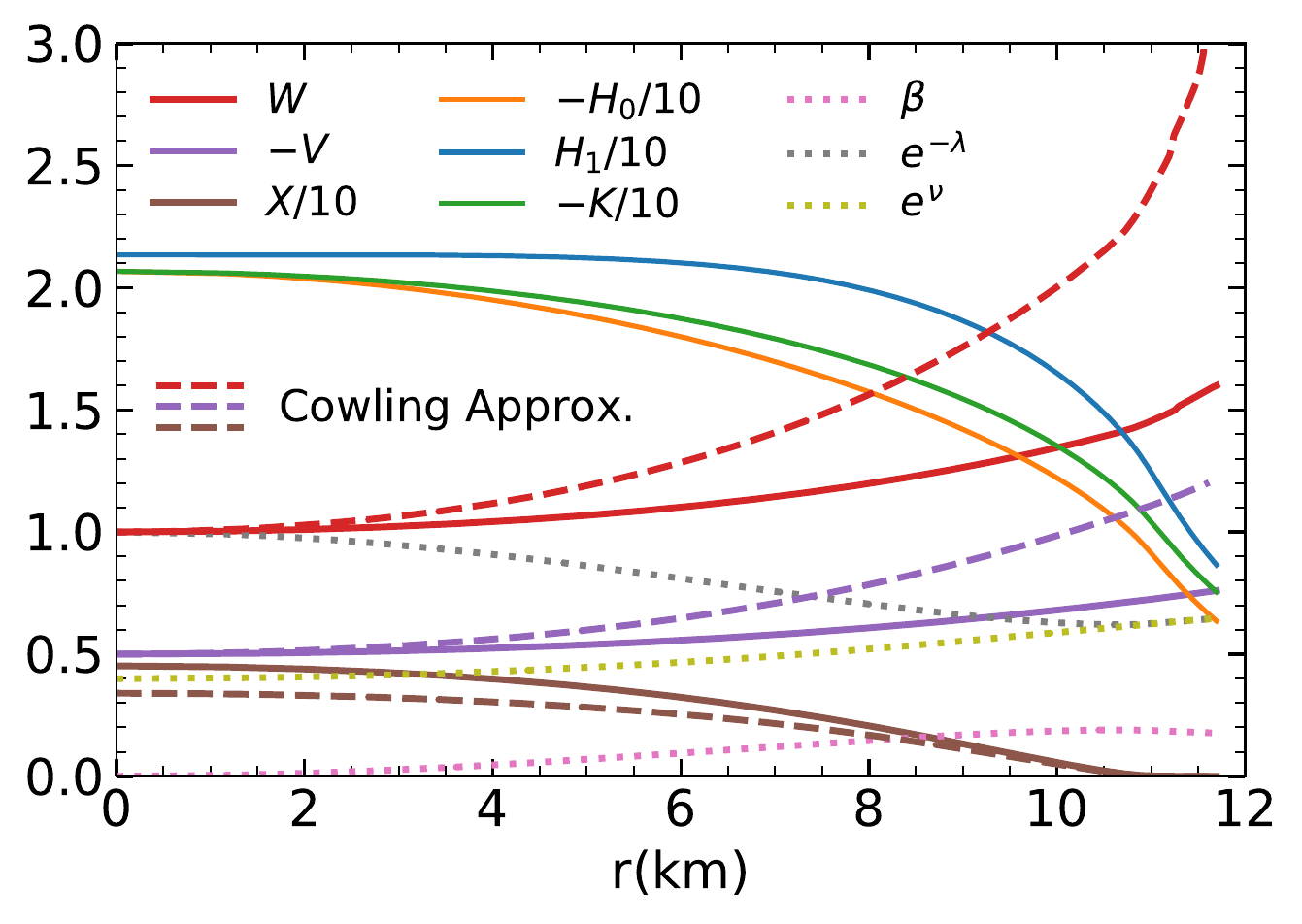}
    \caption{Metric perturbation amplitudes, fluid perturbation amplitudes for non-radial oscillations with $\ell=2$ with (dashed curves) and without (solid curves) the Cowling approximation, and static metric functions (dotted curves) inside a $1.4M_\odot$ NS computed with the Sly4 EOS \cite{Chabanat98}.  $H_0$, $H_1$ and $K$ are in units of $\varepsilon_s=152.26$ MeV fm$^{-3}$, $X$ is in units of $\varepsilon_s^2$, and $W$, $V$, $\nu$ and $\lambda$ are dimensionless. Only real parts of the perturbation amplitudes are plotted. }
    \label{fig:HKWX_EOS_SLY4_1.4}
\end{figure}

As an example, Fig. \ref{fig:HKWX_EOS_SLY4_1.4} shows the static and metric perturbation amplitudes inside a $1.4M_\odot$ NS using the SLy4 EOS.  The central pressure is $p_c=85$ MeV fm$^{-3}$, the radius is $R_{1.4}=11.72$ km, and the corresponding f-mode frequency is $\omega=1.2146\times 10^4+5.206i$ s$^{-1}$, determined as in Sec. \ref{subsec:determine_frequency}. Since the imaginary part of the perturbation is very small compared with the real part, Fig. \ref{fig:HKWX_EOS_SLY4_1.4} shows only the real parts of the perturbation amplitudes. The Cowling approximation gives $\omega=1.5130\times 10^4$ s$^{-1}$, about 25\% larger.  The fluid perturbation amplitudes with the Cowling approximation are in error by up to a factor of 2 at stellar surface for $W$ and $V$, and by about 25\% at the center for $X$, as shown in Fig. \ref{fig:HKWX_EOS_SLY4_1.4}.

Outside the NS, Eq. (\ref{eq:ODE_DL3})  reduces to 2 first-order equations for $H_1$ and $K$; $W$ and $V$ are not defined in this region. These equations can be reformulated into a single Schrodinger-like equation known as the Zerilli equation,
\begin{eqnarray}
{d^2Z}/{dr^{*2}}&=&(V_Z(r)-\omega^2)Z, \label{eq:zerilli}
\end{eqnarray}
by defining \cite{fackerell1971solutions}
\begin{eqnarray}
\begin{pmatrix}
K(r)\\H_1(r)
\end{pmatrix}
&=&
\begin{pmatrix}
g(r) & 1\\
h(r) & k(r)
\end{pmatrix}
\begin{pmatrix}
{Z(r^*)/r}\\ {dZ(r^*)/dr^*}
\end{pmatrix},\nonumber \\
g(r)&=&\frac{n(n+1)+3n {b} +6 {b} ^2}{(n+3 {b} )}, \\
h(r)&=&\frac{(n-3n {b} -3 {b} ^2)}{(1-2 {b} )(n+3 {b} )},\nonumber \\
k(r)&\equiv&\frac{dr^*}{dr}=\frac{1}{1-2 {b} },\nonumber
\end{eqnarray}
and an effective potential 
\begin{eqnarray}
\hspace*{-.5cm}V_Z(r)=(1-2 {b} ) \frac{2n^2(n+1)+6n^2 {b} +18n {b} ^2+18 {b} ^3}{r^2(n+3 {b} )^2}.
\end{eqnarray}
Note that here $b={GM}/{(c^2r)}$ since $m(r>R)=M$. $H_0$ can be fixed by a simplified form of the last of Eq.(\ref{eq:V_def_in}),
\begin{widetext}
\begin{equation}
H_0=\frac{\left[\omega^2r^2-(n+1) {b}  \right]H_1+\left[n(1-2 {b} )-\omega^2 r^2+ {b} (1-3 {b} )  \right] K}{(1-2 {b} )(3 {b} +n)}\label{eq:H0_def_out}.
\end{equation}
\end{widetext}
\begin{figure*}
    \centering
    \includegraphics[width=\linewidth]{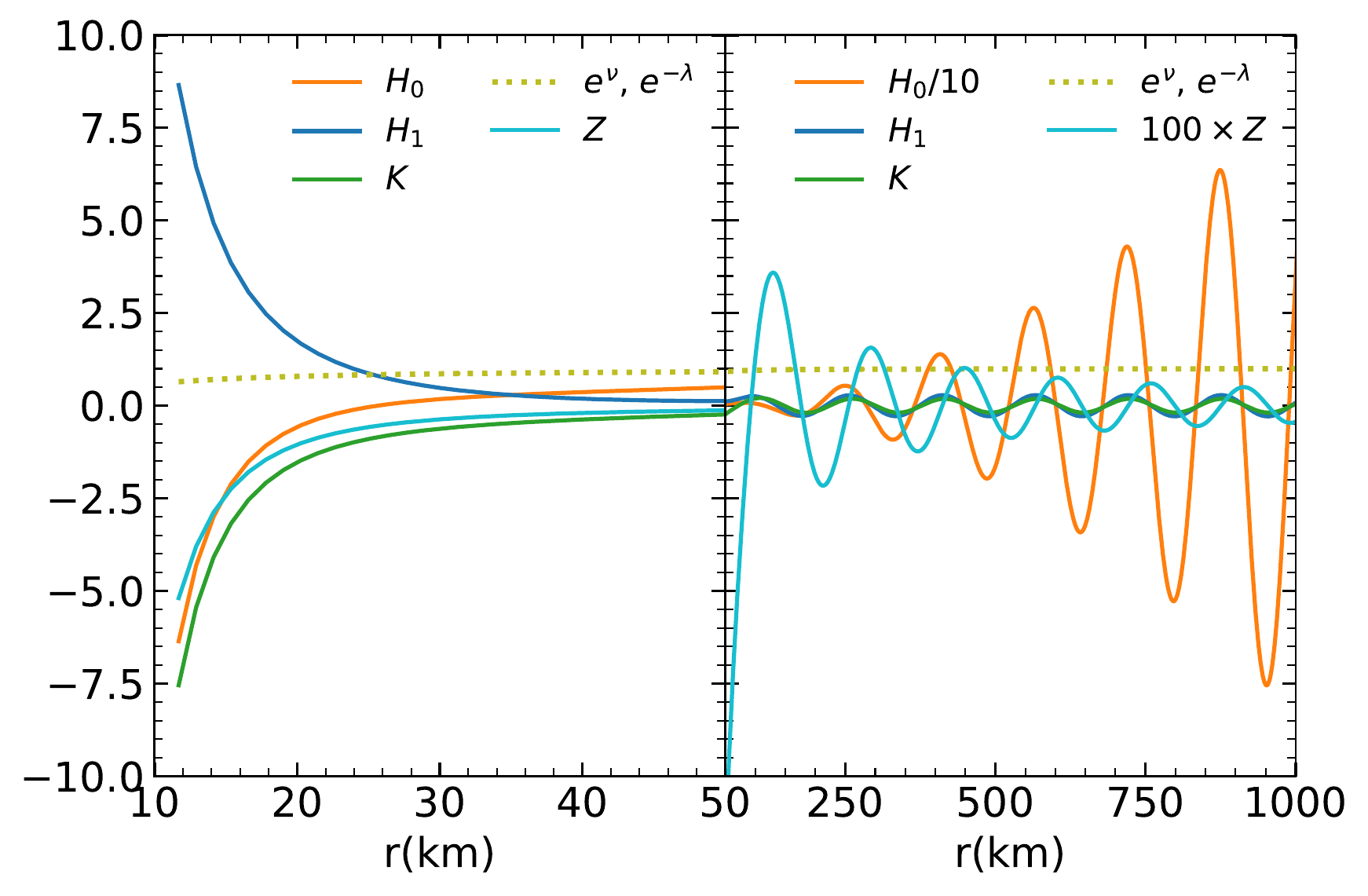}
    \caption{The same as Fig. \ref{fig:HKWX_EOS_SLY4_1.4} except outside the $1.4M_\odot$ SLy4 EOS NS.   Note the change of scale at $r=50$ km.}
    \label{fig:HKZ_EOS_SLY4_1.4}
\end{figure*}
Fig. \ref{fig:HKZ_EOS_SLY4_1.4} shows the metric perturbation amplitudes outside the NS modeled in Fig. \ref{fig:HKWX_EOS_SLY4_1.4}. In the far-field limit, the solution becomes that of oscillating gravitational radiation. 
 The behavior of the metric perturbation amplitudes shown in Figs. \ref{fig:HKWX_EOS_SLY4_1.4} and \ref{fig:HKZ_EOS_SLY4_1.4} is generic for f-mode (n=0) and relatively insensitive to the EOS.
 
\subsection{Determining the oscillation frequency\label{subsec:determine_frequency}}
Our goal is to find the frequencies of eigenmodes that correspond to oscillations. The lowest order mode corresponds to the f-mode generally with zero node, while solutions of higher radial nodes correspond to the g- and p- modes.  All the mode solutions should satisfy the correct boundary condition at infinity should be `free', in other words, at infinity, the gravitational radiation field should be purely outgoing. We solve the Zerilli equation for $r\le25\omega^{-1}$, which is found to be adequate. For larger $r$, the solution of Zerilli equation $Z$ can be decomposed into incoming ($Z_+$) and outgoing ($Z_-$) radiation as
\begin{eqnarray}
\hspace*{-1cm}\begin{pmatrix}
Z(\omega)\\{dZ}/{dr^*}
\end{pmatrix}
&=&
\begin{pmatrix}
Z_-(\omega) \quad Z_+(\omega)\\ {dZ_-}/{dr^*} \quad {dZ_+}/{dr^*}
\end{pmatrix}
\begin{pmatrix}
A_-(\omega)\\A_+(\omega)
\end{pmatrix},\nonumber\\
Z_-&=&e^{-i\omega r^*}\left[\alpha_0+\frac{\alpha_1}{r}+\frac{\alpha_2}{r^2}+\mathcal{O}(r^{-3})\right]\nonumber,\\
\frac{dZ_-}{dr^*}&=&-i\omega e^{-i\omega r^*} [\alpha_0+\frac{\alpha_1}{r} \\
&&+\frac{\alpha_2+{i\alpha_1}(1-2b)/\omega}{r^2}+\mathcal{O}(r^{-3})],\nonumber\\
\alpha_1&=&\frac{-i(n+1)\alpha_0}{\omega},\nonumber\\
\alpha_2&=&\frac{[-n(n+1)+iM\omega({3}/{2}+{3}/{n})]\alpha_0}{2\omega^2},\nonumber
\end{eqnarray}
where $A_+(\omega)$ and $A_-(\omega)$ are the amplitudes of incoming and outgoing radiation, respectively, $Z_+$ is the complex conjugate of $Z_-$.
The amplitude of incoming radiation $A_+(\omega)$ vanishes for physical eigenmodes. $\alpha_0$ can be any complex number which represent an overall phase.

In order to determine $\omega$, we need to solve for the root of $A_+(\omega)=0$ in the complex plane. A straightforward, but inefficient, way would be to use a complex root finding algorithm. With the help of EOS-insensitive relations between $\nu=2\pi$ Re$[\omega]$ and the moment of inertia (see Section \ref{sec:f-I-Lambda}), three digits accuracy can be achieved for the f-mode within 8 trials. Other techniques are needed to guess initial estimates for p- and g-modes.  Note that the imaginary part of the eigenfrequency is usually small ($<1/1000$ the magnitude of the real part) for f-, g- and p-modes. As a result, ${\rm Im}[Z(r^*)]<<{\rm Re}[Z(r^*)]$ as well. Therefore, it's possible to approximately determine the complex eigenfrequency by interpolating $A_+(\omega)$ along the real axis of $\omega$,
\begin{eqnarray}
A_+(\omega) \approx A_0+A_1\omega+A_2\omega^2 =0,
\end{eqnarray}
where $A_0$, $A_1$ and $A_2$ are real constants, by solving the quadratic function $A_+(\omega)$ for real $\omega$ near the eigenfrequency.  This method avoids using a complex root-finding algorithm and is considerably more efficient.

Another simplified method to evaluate eigenmodes is to use the WKB approximation \cite{kokkotas1992w,andersson1995new}. The outside solution is approximated by a WKB wave-function. Perturbation functions near the NS surface can be used to fix amplitude of incoming and outgoing radiation without solving the Zerilli equation. We have verified that these two approximate methods   agreed very well for low-damping modes where the imaginary part is small compared with the real part. Here, we use the interpolation method to determine an initial guess for $\omega$ to be used in the full solution. 

Fig. \ref{fig:beta-f_analytic} shows the f-mode frequency for the SLy4 EOS used in Figs. \ref{fig:HKWX_EOS_SLY4_1.4} and \ref{fig:HKZ_EOS_SLY4_1.4} as a function of neutron star compactness $\beta$.
Note the approximately linear behavior with $\beta^{3/2}$, a universal scaling that becomes apparent when examining analytic results in Newtonican geometry with simple EOSs, as discussed in the next section.

A widely used approximation in the calculation of oscillation frequencies, the Cowling approximation, ignores the metric perturbation $K$, $H_1$ and $H_2$ in Eqs. (\ref{eq:V_def_in}) - (\ref{eq:BC_H1}). This reduces the 4 complex first-order ODEs Eqs. (\ref{eq:ODE_DL3}) to 2 real first-order ODEs and  results in no gravitational radiation damping, see Eqs. (19) and (20) in Ref. \cite{zhao2022quasi}. In addition, in this approximation the Zerilli equation for metric perturbations outside the NS, Eq. (\ref{eq:zerilli}), can be ignored which greatly simplifies the calculation. However, since the Cowling approximation introduces f-mode frequency errors of up to 30\%  \cite{pradhan2022general}, as shown in Fig. \ref{fig:beta-f_analytic}, it is not suitable for the study of the high accuracy universal relations sought in this work.

\section{Newtonian f-mode frequency with analytical EOSs\label{sec:newtonian}}

\begin{figure*}
    \centering
    \includegraphics[width=0.495\linewidth]{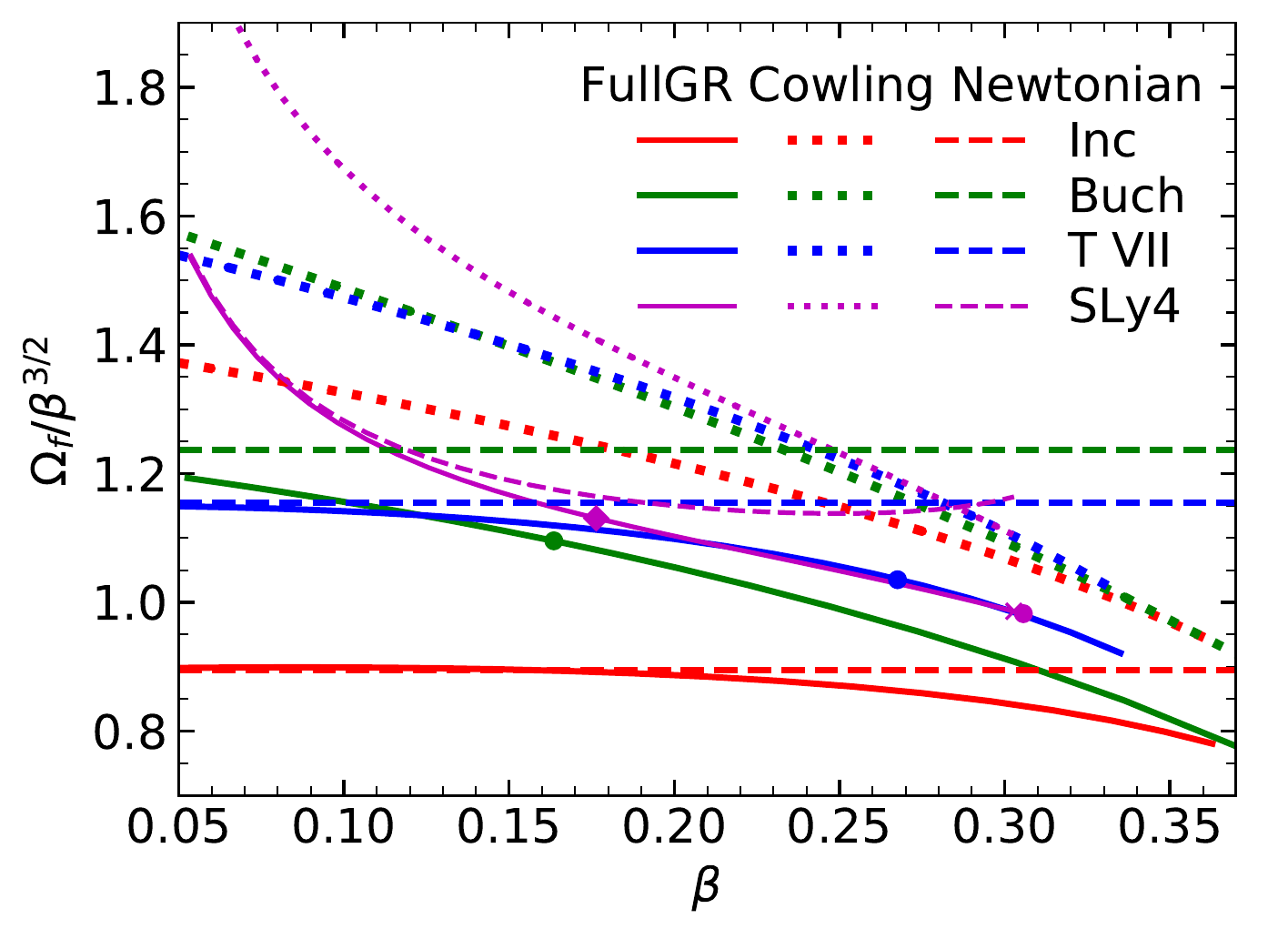}
    \includegraphics[width=0.495\linewidth]{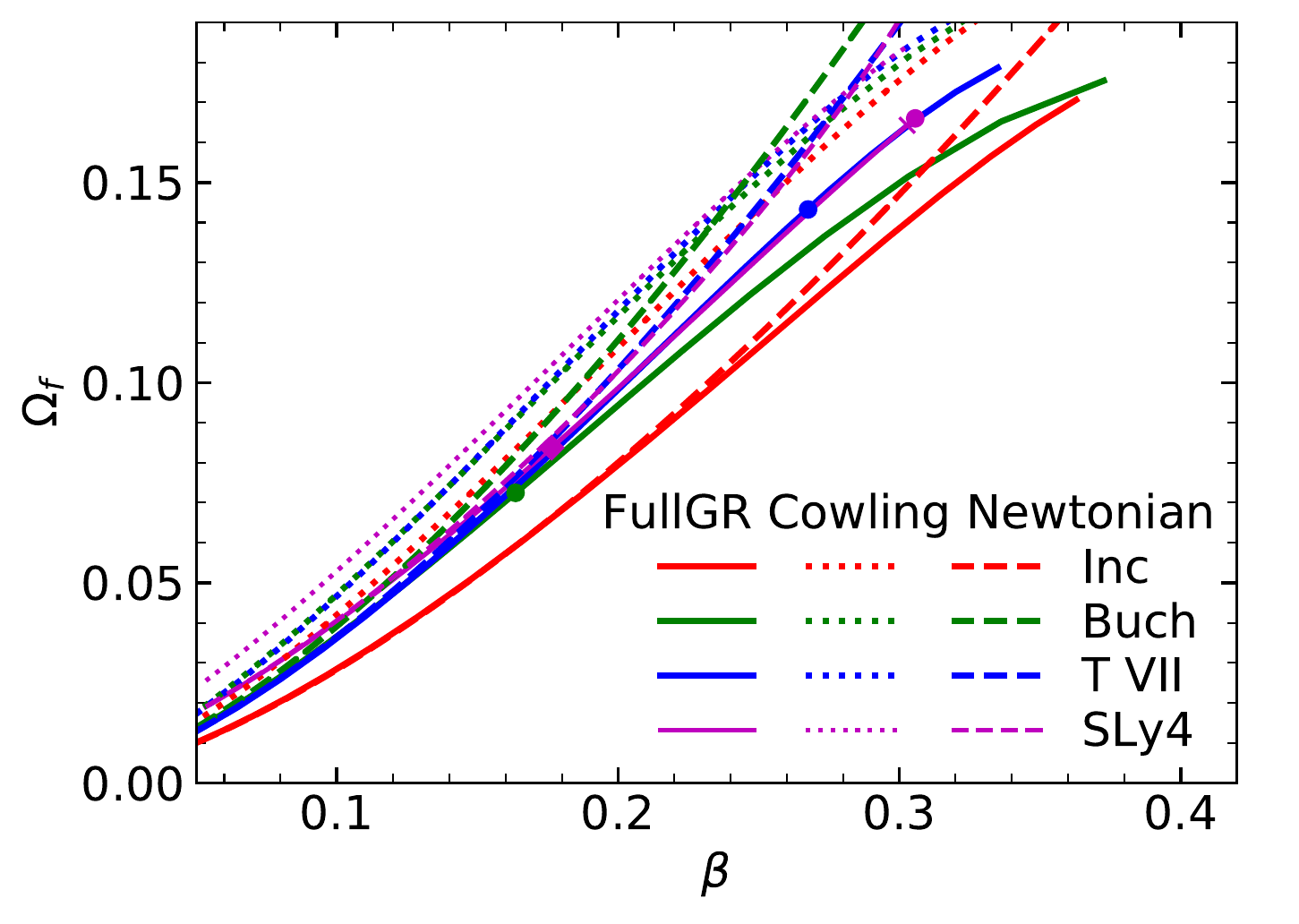}
    \caption[Dimensionless f-mode frequency versus compactness $\beta$ with analytic Newtonian formula and with linear perturbation in general relativity.]{Dimensionless f-mode frequency versus compactness $\beta$ for the SLy4 and three analytic EOSs.  Solid lines show the result of linear perturbation theory in general relativity, dashed lines show the Newtonian result (i.e., the Kelvin-mode frequency), and dotted lines show results assuming the Cowling approximation. Dots on the SLy4, T VII and Buch solid lines indicate $\beta$ values where causality $c_s=c$ occurs; the diamond on the SLy4 solid lines indicates the $1.4M_\odot$ configuration. }
    \label{fig:beta-f_analytic}
\end{figure*}

Much insight can be gained by comparing general relativistic results for realistic EOSs with simplified cases involving analytical EOSs both in general relativity and Newtonian gravity.   In Newtonian physics, a variational analysis on the hydrostatic equilibrium equations with a linear solenoidal velocity perturbation leads to the Kelvin-mode frequency  \cite{chandrasekhar1960general}
\begin{eqnarray}
\omega^2_{Kelvin}&=& G\frac{2\ell(\ell-1)}{2\ell+1} \frac{\int_0^R \varepsilon(r)r^{2\ell-3}m(r) dr}{\int_0^R \varepsilon(r) r^{2\ell} dr}\nonumber\\
&=& G\frac{2\ell(\ell-1)(2\ell-1)}{2\ell+1} \frac{\int_0^R p(r)r^{2\ell-2} dr}{\int_0^R \varepsilon(r) r^{2\ell} dr}
\label{eq:kelvin}\end{eqnarray}
where $\ell$ is the angular quantum number. The second equality in the above uses the hydrostatic equilibrium equations. The Kelvin-mode frequency can be a good approximation to the f-mode frequency of low-mass NSs  with realistic EOSs  \cite{chan2014multipolar}, which we confirm below for $M \lesssim 1.4 M_\odot$ or $\beta\lesssim0.14$. According to Eq. (\ref{eq:kelvin}), Newtonian Kelvin-mode frequencies for the analytic EOSs must satisfy $\omega_{Kelvin}^2=CG{M}/{R^3}$ \cite{andersson1998towards}, or $\Omega_{Kelvin}^2=C\beta^3$, where the coefficients $C$ are displayed in Table \ref{tab:Omega2_over_beta3}.  Fig.  \ref{fig:beta-f_analytic} compares those values to numerical general relativistic results.  A slightly weaker dependence on $\beta$ is apparent for the relativistic results for more massive and compact, but still observable, stars. Because the damping time originates from the general relativistic calculation, and is not defined in the Newtonian calculation, there is no simple analytic estimate for its $\beta$-dependence.

\begin{table}
    \centering
    \caption{Newtonian coefficients $C=\Omega_{Kelvin}^2/\beta^3$  for analytic EOSs}
   \begin{tabular}{|c|c|c|c|}
   \hline
   EOS & $\ell=2$ & $\ell=3$& $\ell=4$ \\\hline
         Inc &  ${4}/{5}$ & ${12}/{7}$ & ${8}/{3}$\\
         ~T VII~ & ${4}/{3}$ & ${204}/{77}$ & ${152}/{39}$\\
         Buch& ~$3\pi^2(5\pi^2-30)^{-1}$~ & ~2.94766~ & ~4.24121~ \\\hline
    \end{tabular}
     \label{tab:Omega2_over_beta3}
\end{table}

In the case of an homogeneous incompressible sphere (Inc), the energy density is constant,  $\varepsilon(r)=\varepsilon_*$, and one finds  \cite{cowling1941non}
\begin{eqnarray}
C&=&\frac{2\ell(\ell-1)}{2\ell+1}.
\end{eqnarray}
This EOS will be seen to be a good approximation to that of pure quark stars.

In the case of the Tolman VII \cite{tolman1939static} (T VII) solution, the energy density has a quadratic dependence, $\epsilon(r) = \varepsilon_c[1-(r/R)^2]$, and
\begin{eqnarray} C&=&\frac{2\ell(\ell-1)(2\ell+11)}{(2\ell+1)(2\ell+5)}.
\end{eqnarray}
This analytic solution is seen to be an excellent approximation to the realistic SLy4 EOS for observable neutron stars in the relativistic case for both f-mode frequencies and damping times.

A third analytical solution is due to Buchdahl \cite{buchdahl1959general} (Buch), and it is the only one stemming from an analytic EOS, $\varepsilon=12\sqrt{p_*p}-5p$ where $p_*$ is a constant. In the Newtonian limit, Buch becomes equivalent to the $n=1$ polytropic EOS $p=\varepsilon^2/(144p_*)$, for which $\varepsilon(r) \propto (R/r)\sin(r/R)$ and 
\begin{eqnarray}
C&=&-\frac{2\ell(\ell-1) {~}_1F_2(\ell-\frac{1}{2};\frac{3}{2},\ell+\frac{1}{2};-\pi^2)}{(2\ell-3) {~}_1F_2(\ell+\frac{1}{2};\frac{3}{2},\ell+\frac{3}{2};-\frac{\pi^2}{4}) },
\end{eqnarray}
where ${}_AF_B$ is the generalized hypergeometric function.  This analytic solution has less success approximating a realistic EOS such as SLy4 than does T VII.

The relativistic Cowling approximation is applied to the three analytic and the SLy4 EOS, see dotted lines in Fig. \ref{fig:beta-f_analytic}, and is seen to generally overestimate $\Omega_f$ by 20-30\%. In case of Inc, the deviations  exceed 40\% at low compactness where the Newtonian result is extremely accurate. Even in the SLy4 case, the Newtonian estimation is much more reliable except for extreme compactness where the error introduced by static Newtonian gravity dominates over that due to ignoring gravitational perturbations.

\section{EOS for hadronic, hybrid and pure quark stars \label{sec:EOS}}

We are interested in hadronic, pure quark and hybrid NSs with first-order phase transition. Previous studies of NS f-and g-modes with first-order phase transitions used polytropic EOSs with the same polytropic index for both low- and high-density parts of EOS  \cite{sotani2001density,miniutti2003non} or various hadronic EOSs with the MIT bag model \cite{yip1999quadrupole,sotani2011signatures,flores2014discriminating}. An improvement we seek is  to calculate with a more general hadronic NS EOS constrained by \chiEFT N3LO calculations \cite{drischler2020limiting} and causality coupled, if necessary, to constant sound speed matter at densities above a first-order phase transition.

The well-understood outer crust EOS is dominated by relativistic degenerate electron with pressure $p_e= ({3}/{\pi^2})^{1/3} n_e^{4/3}/4$ with a small (negative) contribution from the ionic lattice. Dripped neutrons appear in the NS inner crust and slightly further modify the pressure.  The hadronic contributions cause no more than a 10\% deviation from $p_e$. Uncertainties in the nuclear interaction in the crust have a relatively negligible effect compared with uncertainties in the core EOS. As a result we use the same fixed crust EOS for all hadronic and hybrid NSs in this study. We employ an analytic crust EOS, which is presented in Appendix \ref{sec:crustEOS}, to avoid interpolation errors and to speed up the eigenvalue solver, so that a multitude of models may be considered. We have ensured this simplification introduces a negligible error (Fig. \ref{fig:crust_eos_compare}). 

Read et al.~ \cite{read2009constraints} found the pressure $p$ of realistic high-density cold matter could be relatively faithfully rendered using three polytropic segments, each segment being described by $p=K_in^{\gamma_i}$ within the region $n_{i-1}<n<n_i$ where $ i=1-3$. The matching condition at the core-crust transition together with continuity of both pressure $p$ and energy density $\varepsilon$ at the boundaries determine $K_i$, leaving 6 free parameters, $n_i$ and $\gamma_i$ where $i\in=[1-3]$.  Equivalently, $n_i$ and $p_i$ can be used as parameters. Within the polytropic segment $i$, the energy density is given by
\begin{equation}
\varepsilon=\varepsilon_{i-1}\frac{n}{ n_{i-1}}+\frac{p-p_{i-1}(n/n_{i-1})}{\gamma_i-1},\quad n_{i-1}\le n\le n_i.
\end{equation}
The polytropic exponents and the energy densities at the boundaries are given by
\begin{eqnarray}
\quad\varepsilon_i&=&\frac{p_i}{\gamma_i-1}+\left(\varepsilon_{i-1}-\frac{p_{i-1}}{\gamma_i-1}\right)\frac{n_i}{ n_{i-1}}, \cr
\gamma_i&=&\frac{\ln(p_i/p_{i-1})}{\ln(n_i/n_{i-1})}.
\end{eqnarray}

We take $n_0=0.04$ fm$^{-3}$, $\varepsilon_0=37.88$ MeV fm$^{-3}$, $p_0=0.1239$ MeV fm$^{-3}$ from the crust EOS of SLy4 (  see Appendix A).   Following Read et al., we choose $n_1=0.3$ fm$^{-3}$, $n_2=0.6$ fm$^{-3}$, $n_3=1.2$ fm$^{-3}$ \cite{read2009constraints}. The polytrope for $n_0<n<n_1$ is fitted to $\pm \sigma$ ($\pm 2\sigma$) NS matter EOS with a $\chi$EFT N3LO calculation \cite{drischler2020limiting}, giving $p_1= 15.63\pm 3.54 (\pm7.09)$ MeV fm$^{-3}$. The EOS in the higher density region $n>n_1$ is controlled by $p_2$ and $p_3$, which are free parameters limited by causality and maximum neutron star mass constraints \cite{zhao2018tidal}. We generate about 3000 hadronic EOSs by exploring the entire allowed ranges of $p_1$, $p_2$ and $p_3$. This 3-parameter model is called \emph{PP3}.

For the high-density matter above the phase transition in hybrid stars, or for strange quark stars, we use a constant sound speed  (called \emph{CSS})  \cite{alford2013generic} EOS. 
In the high-density matter with $p>p_t$ in hybrid stars, one has
\begin{equation}
\varepsilon=(p-p_{t})/c_{s}^2+\varepsilon_{t}+\Delta\varepsilon
\end{equation}
where $p_t$ is the pressure at the first-order phase transition, $\varepsilon_t$ is the energy density at the low-density side of the phase transition, $\Delta\varepsilon$ is the density jump at the phase transition, and $c_s$ is the (constant) sound speed for $p>p_t$.

There are 3 parameters for the \emph{CSS} part of the hybrid EOS; $n_t/n_s\in[1,4]$, $\Delta \varepsilon/\varepsilon_t\in[0.1,1]$, and $c_s^2\in[1/3,1]$. In order to minimize parameters in the hadronic EOS for hybrid stars, we only vary $p_1$ within the 2$\sigma$ \chiEFT band and fix $p_2=7.3 p_1$ and $p_3=7.3 p_2$, so that $\gamma_2=\gamma_3=2.867$ $\gamma_1=\gamma_2=2.868$ for $n>n_1$ whenever needed. We therefore utilize only 3 hadronic EOSs, which we call N3LO-cen and N3LO$\pm\sigma$, for hybrid stars. Note that this is not the same as using the published N3LO EOS, which is only given up to about $2n_s$, but employs a particular extrapolation for $n>2n_s$ as described in Ref. \cite{drischler2020limiting}.  In total, there are 4 parameters for the hybrid stars.

In pure quark stars at all densities
\begin{equation}
\varepsilon=p/c_{s}^2+\varepsilon_{surf}
\end{equation}
where $\varepsilon_{surf}$ is the energy density at the surface (where $p=0$). The quark and hybrid expressions are the same when $p_t=\Delta\varepsilon=0$ and $\varepsilon_t=\varepsilon_{surf}$. For the pure quark EOS, there are only two parameters, $c_s$ and $\varepsilon_{surf}$.

\section{f-mode frequencies of hadronic, hybrid and quark stars}

\begin{figure*}
    \centering
    \includegraphics[width=\linewidth]{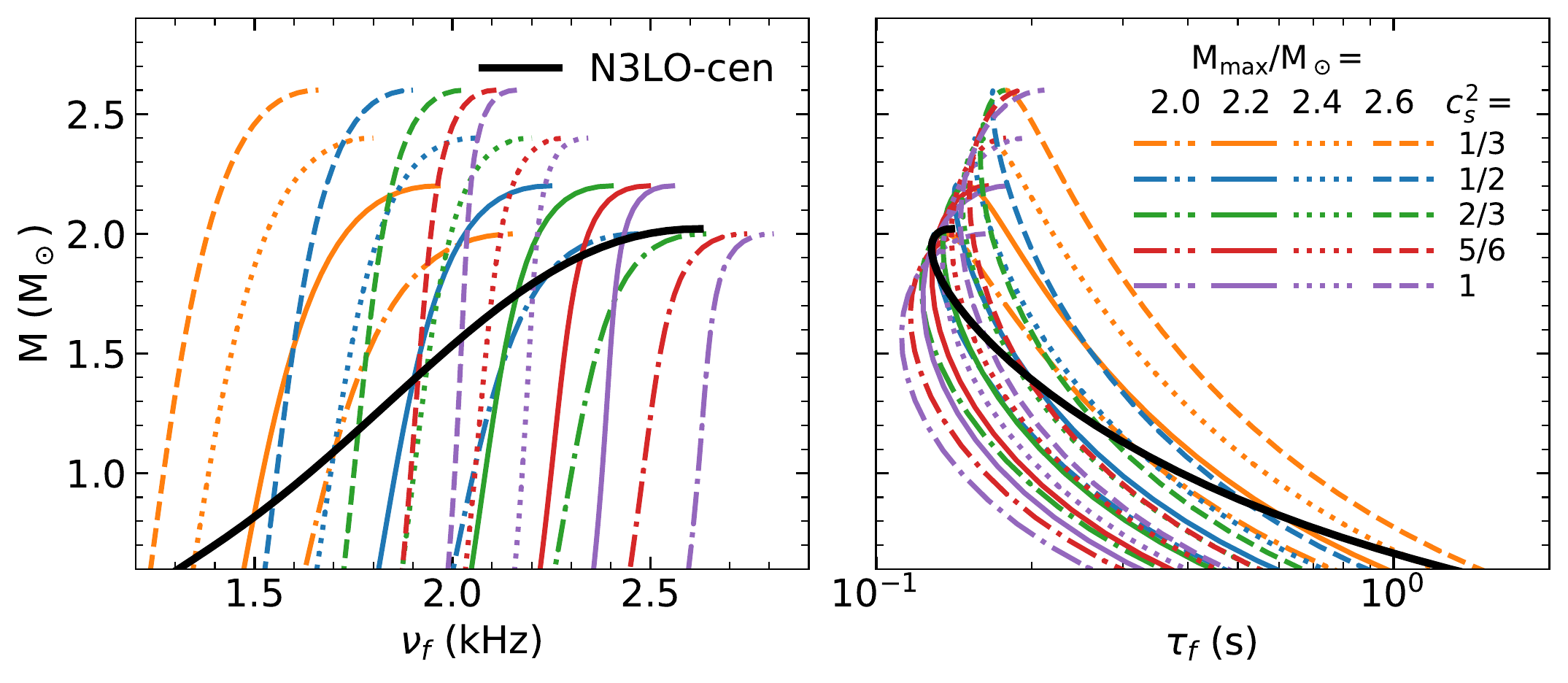}
    \caption[f-mode frequency and damping time versus mass for pure quark stars with \emph{CSS}]{The left (right) panel shows the f-mode frequency (damping time) as a function of $M$ for pure quark stars with the CSS parameterization. \label{fig:f-mode_strange}}
\end{figure*}
We first compare f-mode frequencies as a function of stellar mass for pure quark stars (i.e., self-bound stars) using the \emph{CSS} parameterization with those for representative hadronic EOSs corresponding to the central values of $\chi$EFT neutron matter calculation (N3LO-cen) and its one-sigma uncertainty bounds (N3LO$\pm\sigma$) for all densities in excess of the crust-core boundary.  In this example, $c_s^2$ ranges from 1/3 to 1 while the other parameter $\varepsilon_{surf}$ is selected to give a maximum neutron star mass $M_{\rm max}$ from 2.0 to 2.6 $M_\odot$.
Note that quark stars with smaller $M_{\rm max}$ have larger f-mode frequencies, and that in all cases, the frequency increases with stellar mass.  Quark stars also have a lower limiting frequency whereas the hadronic frequencies decrease continuously with decreasing mass to low frequency and mass.  Damping times for quark stars, on the other hand, generally decrease with mass except near $M_{\rm  max}$ and there is no threshold damping time. 

\begin{figure*}
    \centering
    \includegraphics[width=\linewidth]{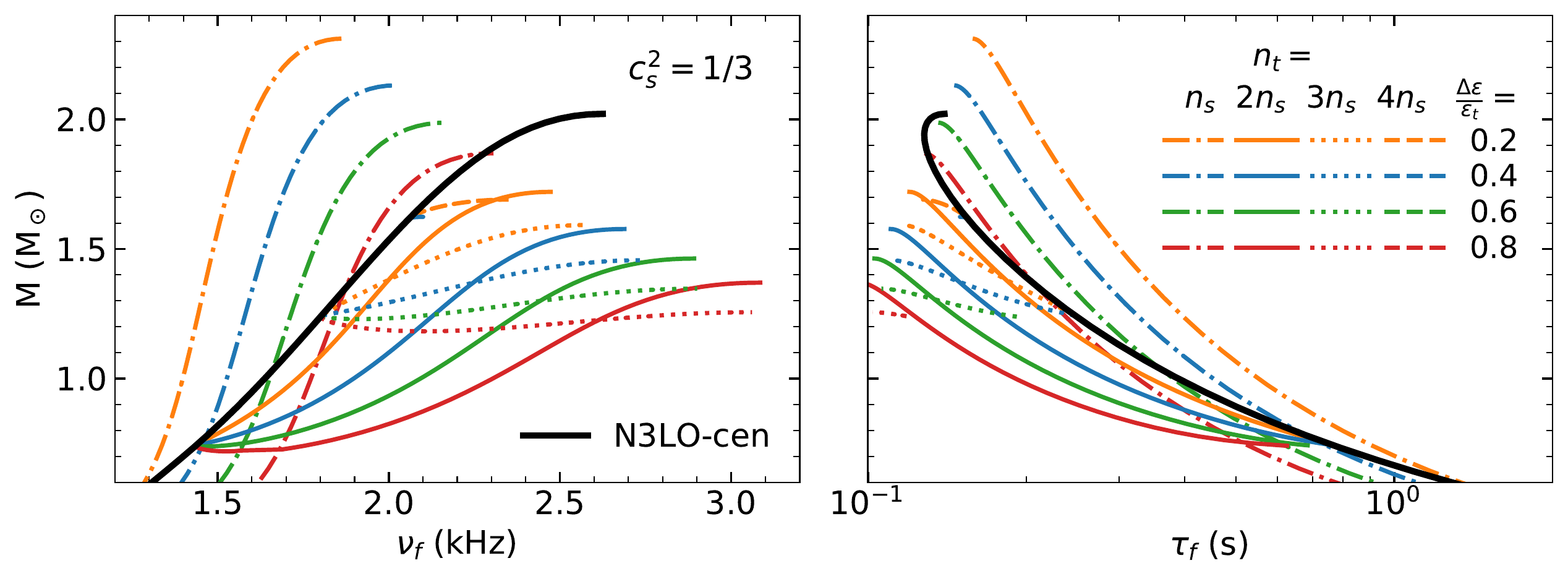}
    \includegraphics[width=\linewidth]{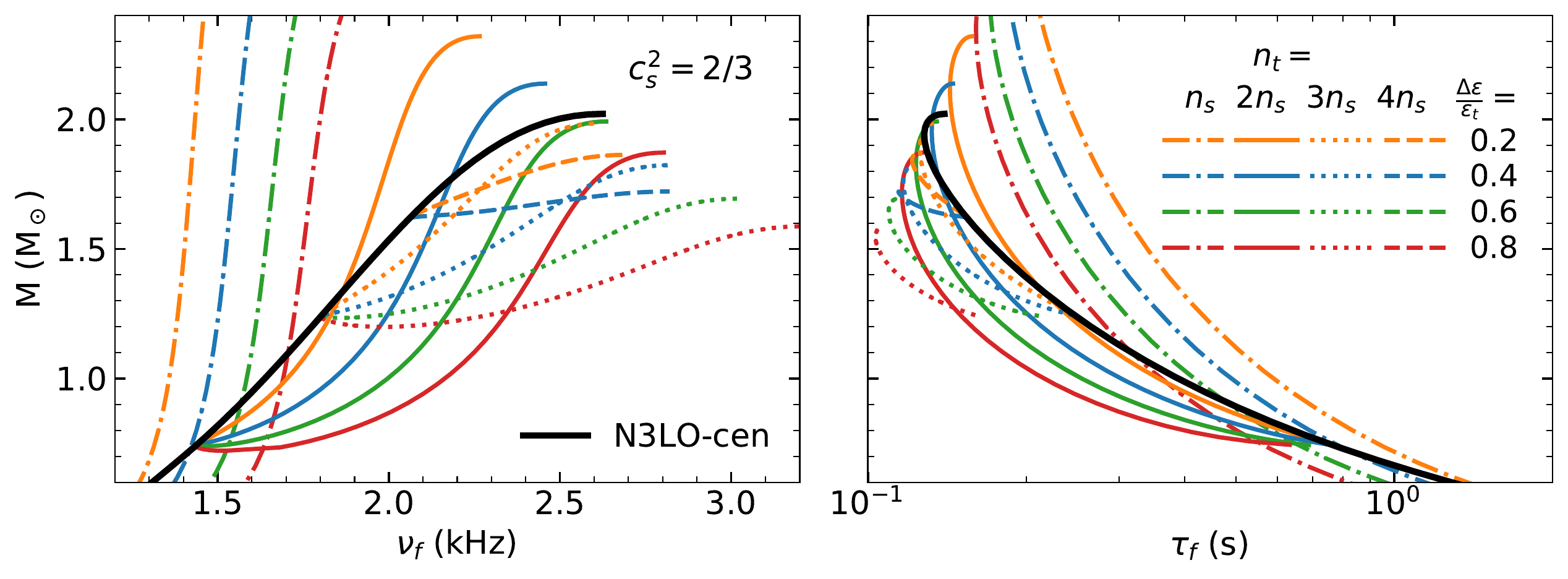}
    \includegraphics[width=\linewidth]{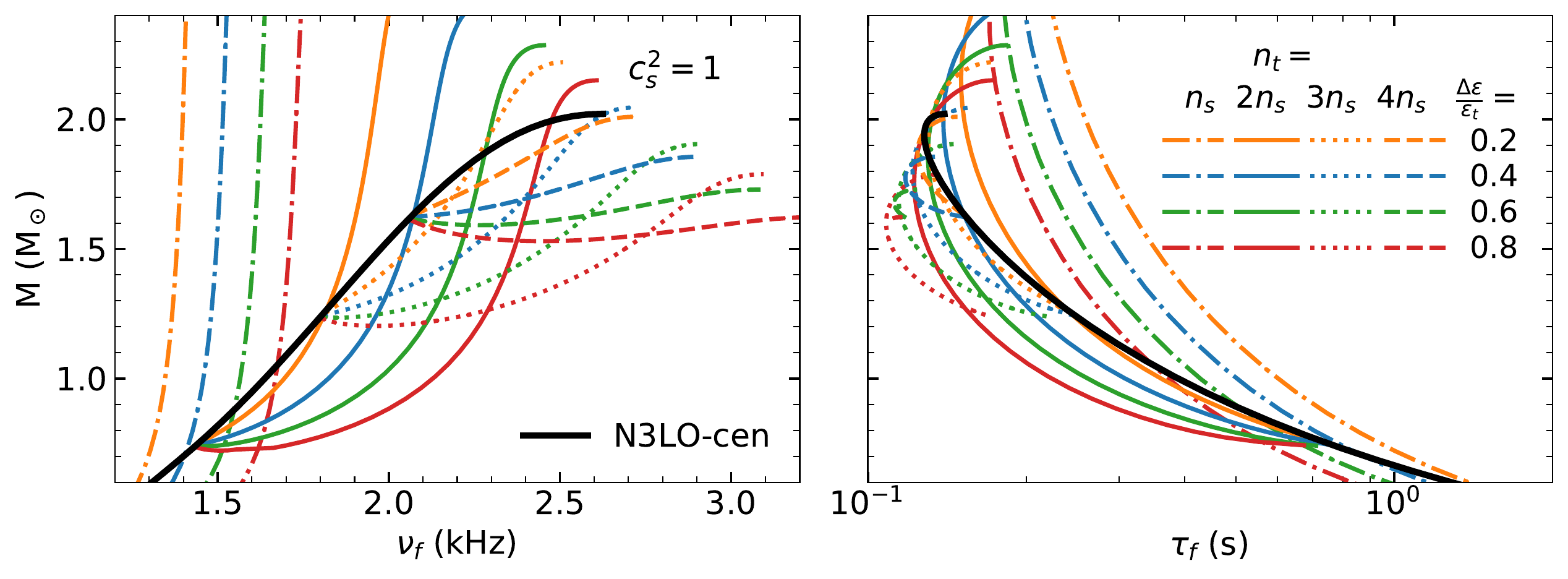}
\caption[f-mode frequency and damping time versus mass for hybrid NS with $\chi$EFT and \emph{CSS}]{The same as Fig. \ref{fig:f-mode_strange} but for hybrid NSs with the hadronic EOS fitted to N3LO-cen and \emph{CSS} sound speeds $c_s^2=[1/3, 2/3, 1]$ from top to bottom, respectively.\label{fig:f-mode_hybrid}}
\end{figure*}
 Next we explore the behavior of hybrid stars in Fig. \ref{fig:f-mode_hybrid}. Various values of $n_t$, $\Delta\varepsilon/\varepsilon_t$ and $c_s^2$ are utilized. No restrictions are placed on $M_{\rm max}$ for this study. As in Fig. \ref{fig:f-mode_strange}, the hadronic EOS N3LO-cen is used in the hadronic parts of the hybrid stars.
For $n_t=n_s$, the phase transition occurs for masses so small that the resulting f-mode frequencies and damping times show a similar behavior to those of pure quark stars.
Just above the transition masses, the frequencies increase more rapidly with mass for hybrid stars than for the N3LO-cen hadronic stars, but this behavior can reverse close to $M_{\rm max}$.

\begin{figure*}
    \centering
    \includegraphics[width=0.495\linewidth]{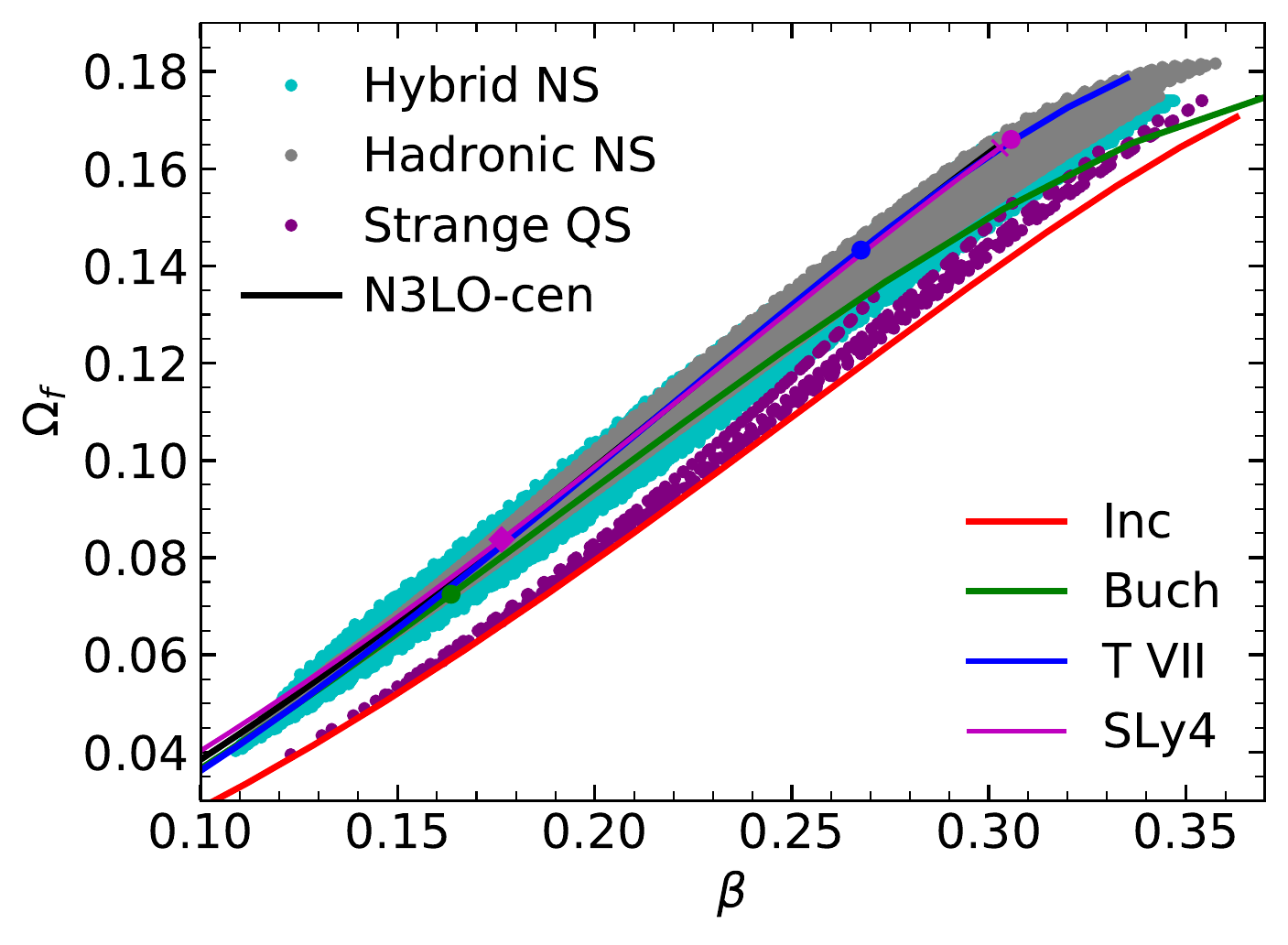}
    \includegraphics[width=0.495\linewidth]{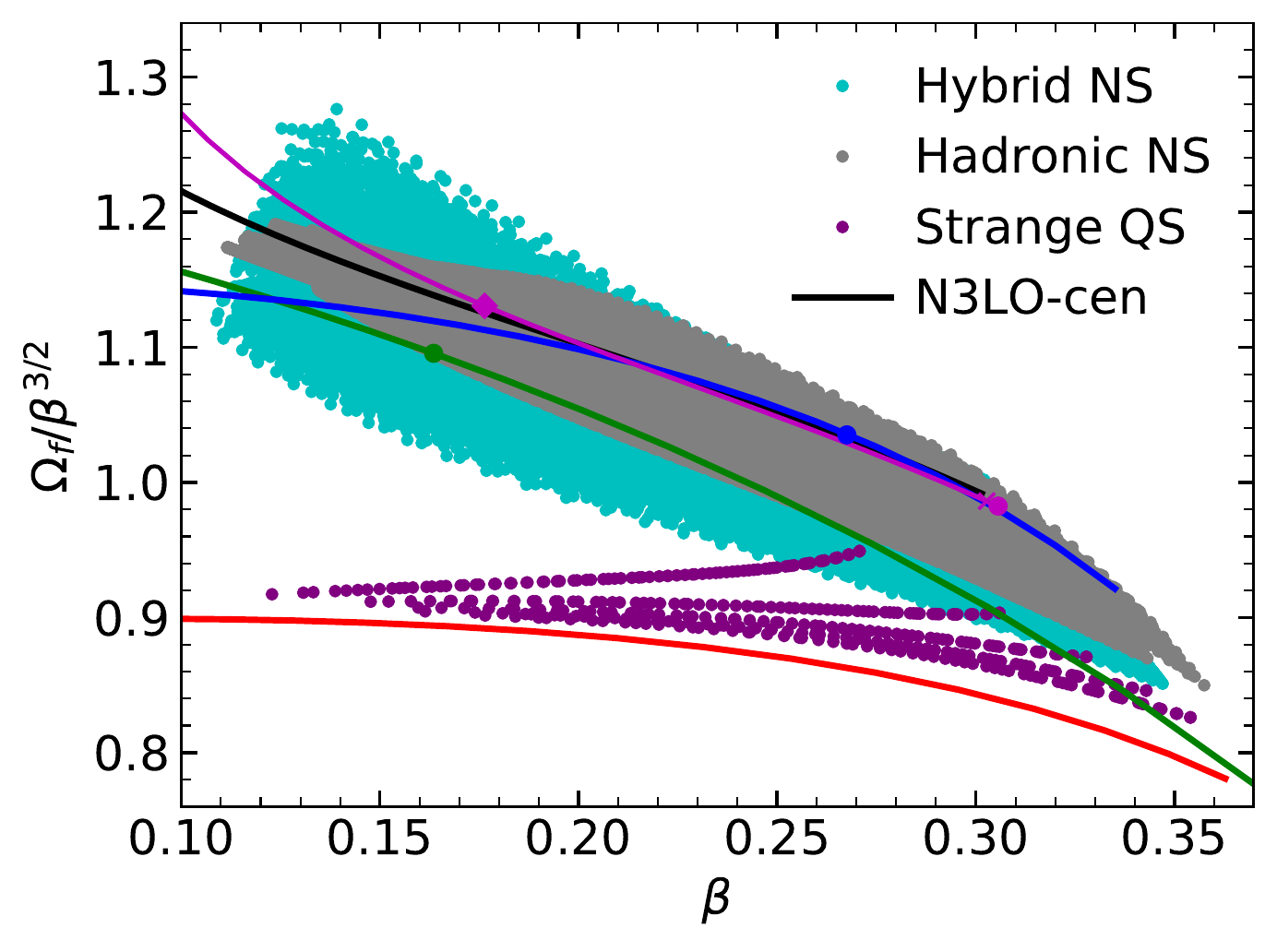}
    \caption[Dimensionless frequency versus compactness $\beta$ with analytic Newtonian formula and with linear perturbation in general relativity.]{The dimensionless f-mode frequency $\Omega_f$ as a function of stellar compactness $\beta$ for parameterized hadronic, hybrid and quark matter stars constained by causality and $M_{\rm max}\ge2M_\odot$. General relativistic results for the analytic EOSs and the SLy4 and N3LO-cen hadronic EOSs are shown for comparison.}
    \label{fig:beta-f}
\end{figure*}
Next, we proceed to an examination comparing larger numbers of hadronic, hybrid and quark matter stars in order to identify systematic trends.   Fig. \ref{fig:beta-f} shows the relation between the f-mode frequency and the stellar compactness for thousands of stars of varying mass:  hadronic stars having a fixed crust with the 
\emph{PP3} parameterized EOS, hybrid stars having a four-parameter EOS with a fixed crust, and the three parameter \emph{CSS} quark matter stars, all subject to the constraints of causality and $M_{\rm max}\ge2M_\odot$. As expected from our earlier comparison of Sly4 and the three analtyic EOSs, there is a global correlation between $\Omega_f$ and $
\beta$.  However, in contrast to the Newtonian analytic EOSs, the purely linear relation between $
\Omega_f$ and $\beta^{3/2}$ is broken.
The right-hand panel in Fig. \ref{fig:beta-f} shows that the effective power of $\beta$ that would best fit quark stars remains close to 3/2, but best fits to the body of hadronic and hybrid stars is considerably less than 3/2.  

\begin{figure*}
    \centering
    \includegraphics[width=0.495\linewidth]{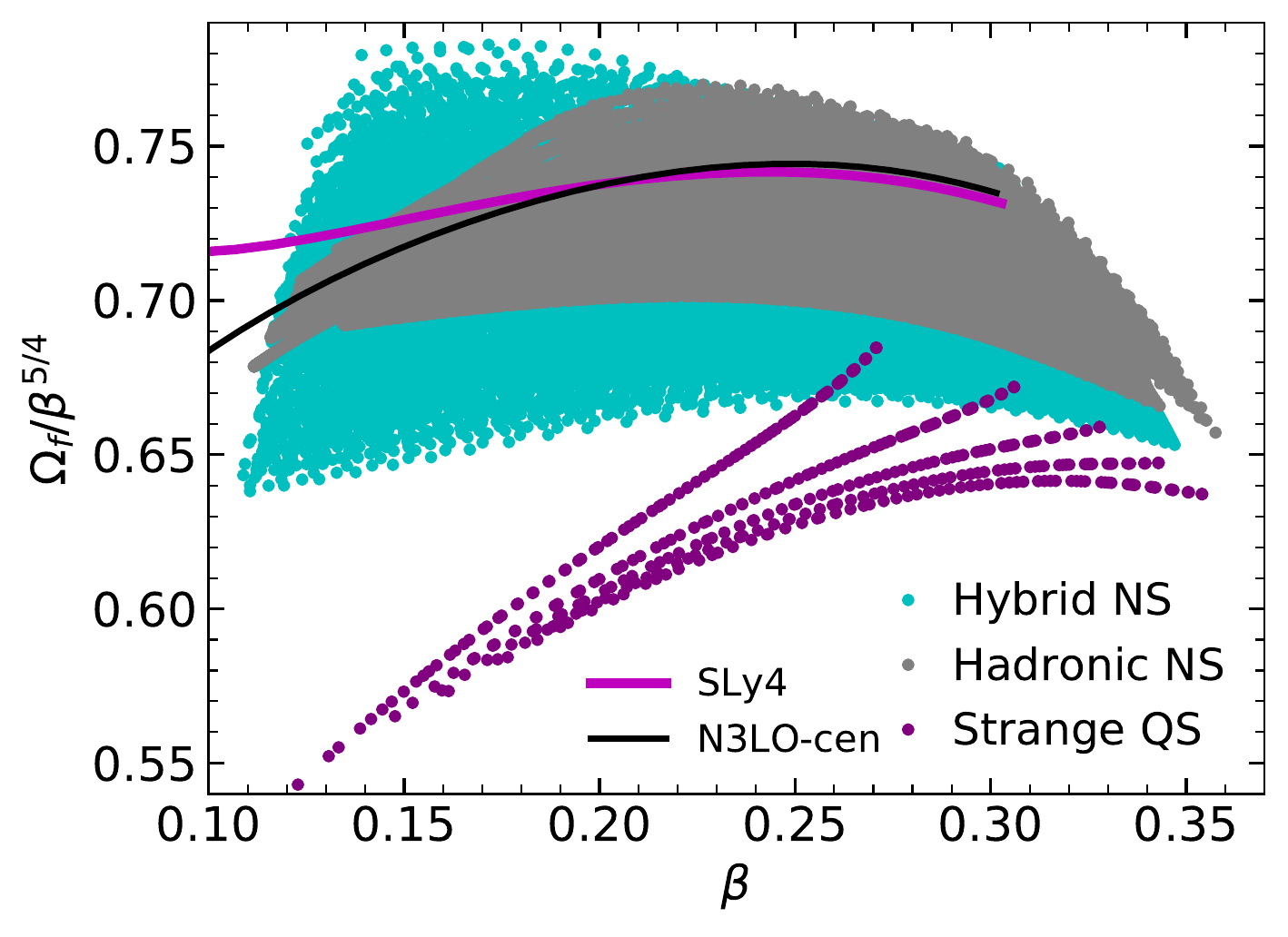}
    \includegraphics[width=0.495\linewidth]{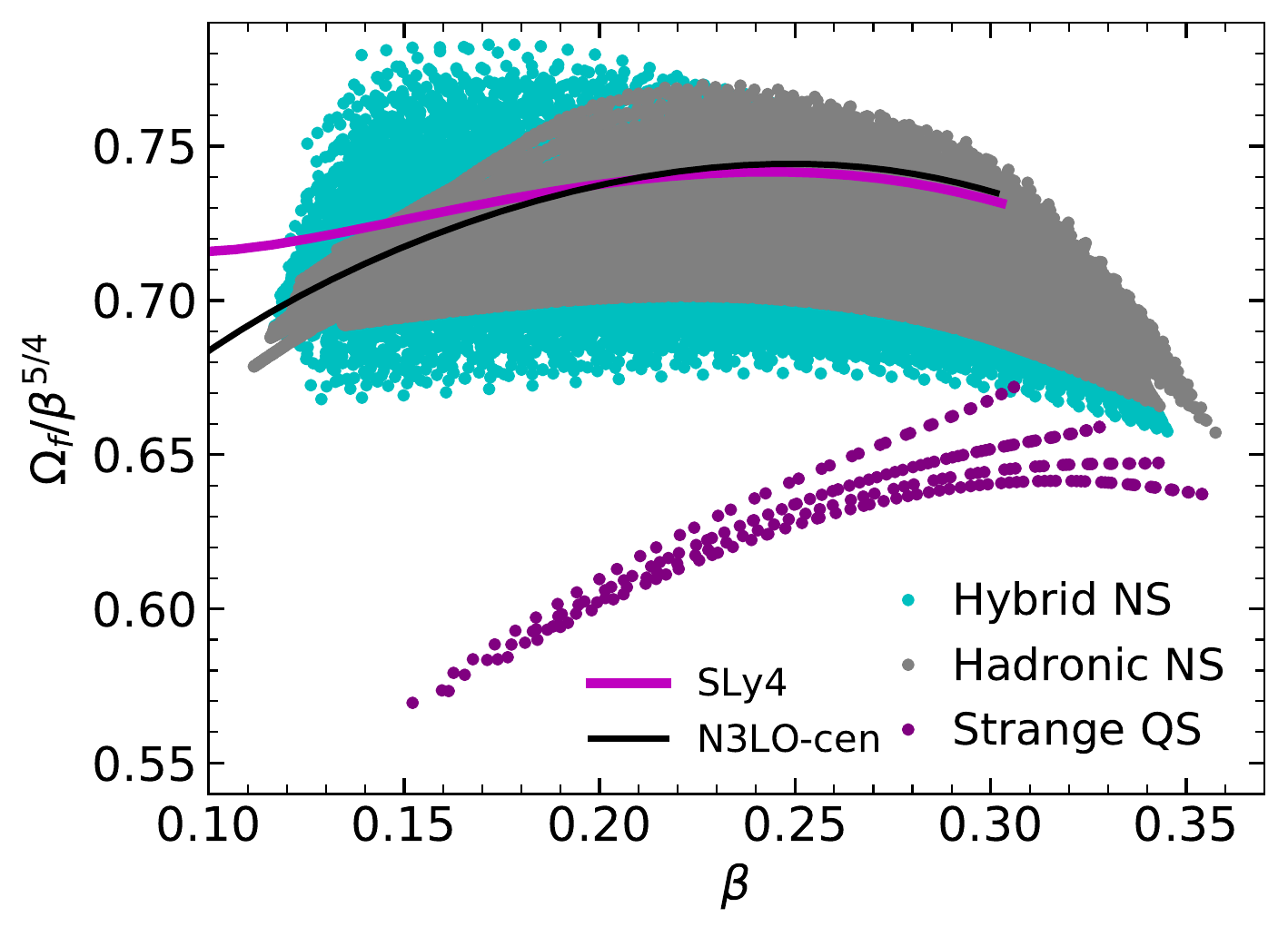}
    \caption[]{Left: the same as the right panel of Fig. \ref{fig:beta-f} but with the y-axis replaced by $\Omega_f \beta^{5/4}$. Right: the same as the left panel but including the additional GW170817 $\tilde\Lambda$ constraint \cite{de2018tidal}.}
    \label{fig:beta-f1}
\end{figure*}

We find a power-law fit with the minimum uncertainty, applicable to stars with $M\ge1M_\odot$, is
\begin{equation}
    \Omega_f=A\beta^{5/4},
    \label{eq:Obfit}
\end{equation}
where $A=0.714\pm0.056$ for hadronic stars and $A=0.711\pm0.072$ for hybrid stars (see the left panel of Fig. \ref{fig:beta-f1}).  If we further impose the GW170817 binary tidal deformability constraint $\bar \Lambda_{GW170817}<521$ \cite{de2018tidal}, the respective values of $A$ stay the same for hadronic stars and $A=0.720\pm0.063$ for hybrid stars (see the right panel of Fig. \ref{fig:beta-f1}).

\section{$\Omega_f-\bar I-\Lambda$ Relation\label{sec:f-I-Lambda}}
Although we found a simple power-law fit relating $\Omega_f$ to $\beta^{5/4}$ for all hadronic and hybrid stars that has only an 8\% uncertainty, more powerful correlations exist that relate $\Omega_f$ to $\bar I$ and $\Lambda$.   It has already been demonstrated that the universal relations relating the f-mode to moment of inertia and tidal deformability are much tighter than similar relations involving other modes, e.g. $p$-modes \cite{andersson1998towards}, and $w$-modes  \cite{benitez2021investigating}.

\begin{figure*}
    \centering
    \includegraphics[width=\linewidth]{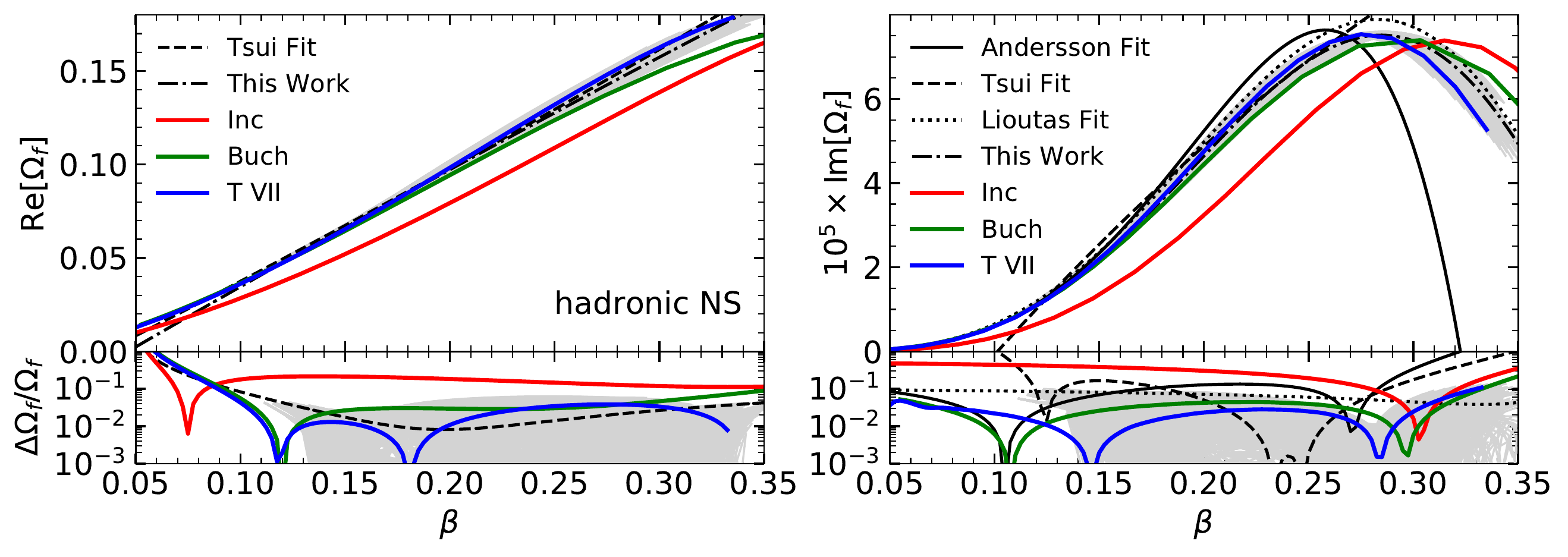}
    \includegraphics[width=\linewidth]{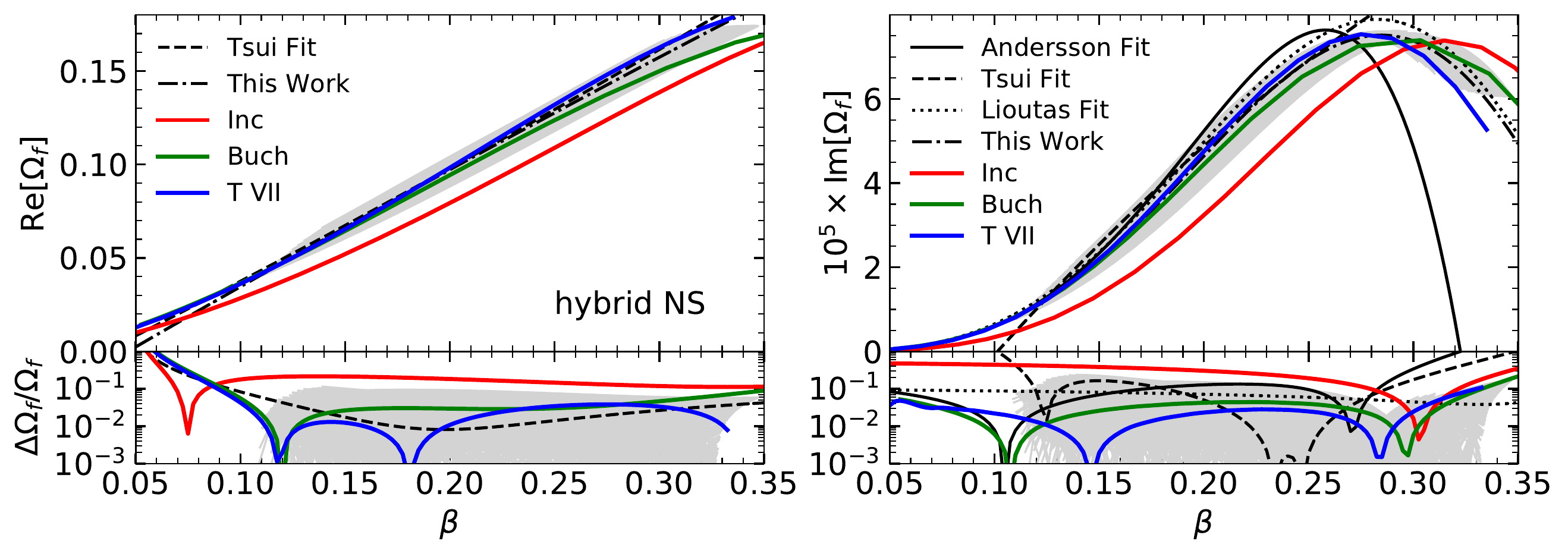}
    \includegraphics[width=\linewidth]{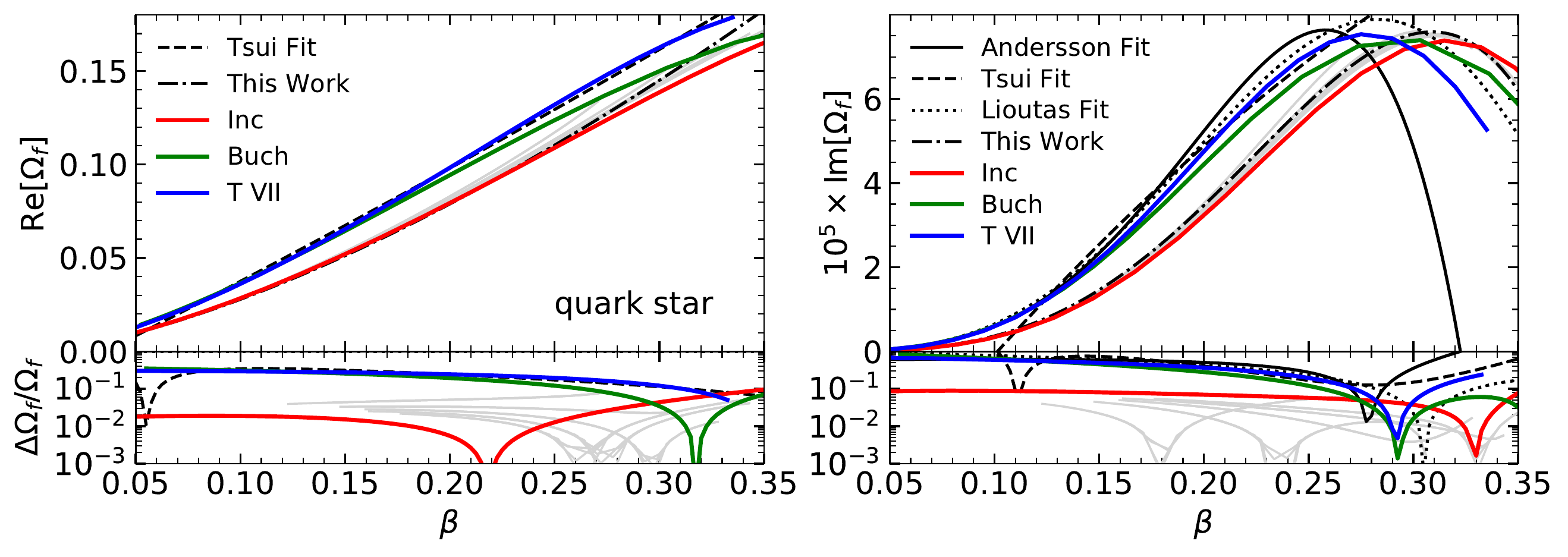}
    \caption[$f-\beta$ EOS insensitive relation]{The $\Omega_f-\beta$ EOS insensitive relations. The left (right) panel shows the real (imaginary) part of the dimensionless frequency. The top, middle, and the bottom panels are for hadronic, hybrid, pure quark (self-bound) EOSs, respectively.  ``This Work'' refers to the fits of Eq. (\ref{eq:fbfit}) for hadronic and hybrid stars and Eq. \ref{eq:fbfitstrange} for pure quark stars.   Fits due to Refs.  \cite{tsui2005universality} (Tsui), \cite{andersson1998towards} (Andersson), and  \cite{lioutas2021frequency} (Lioutas) are also shown.  The lower parts of each panel show residuals of fits from this work; deviations from this work of the relativistic numerical results for individual parameterized EOSs are shown as gray lines (which blend into gray regions for hadronic and hybrid stars due to their density).}
    \label{fig:omega-beta}
\end{figure*}
\begin{figure*}
    \centering
    \includegraphics[width=\linewidth]{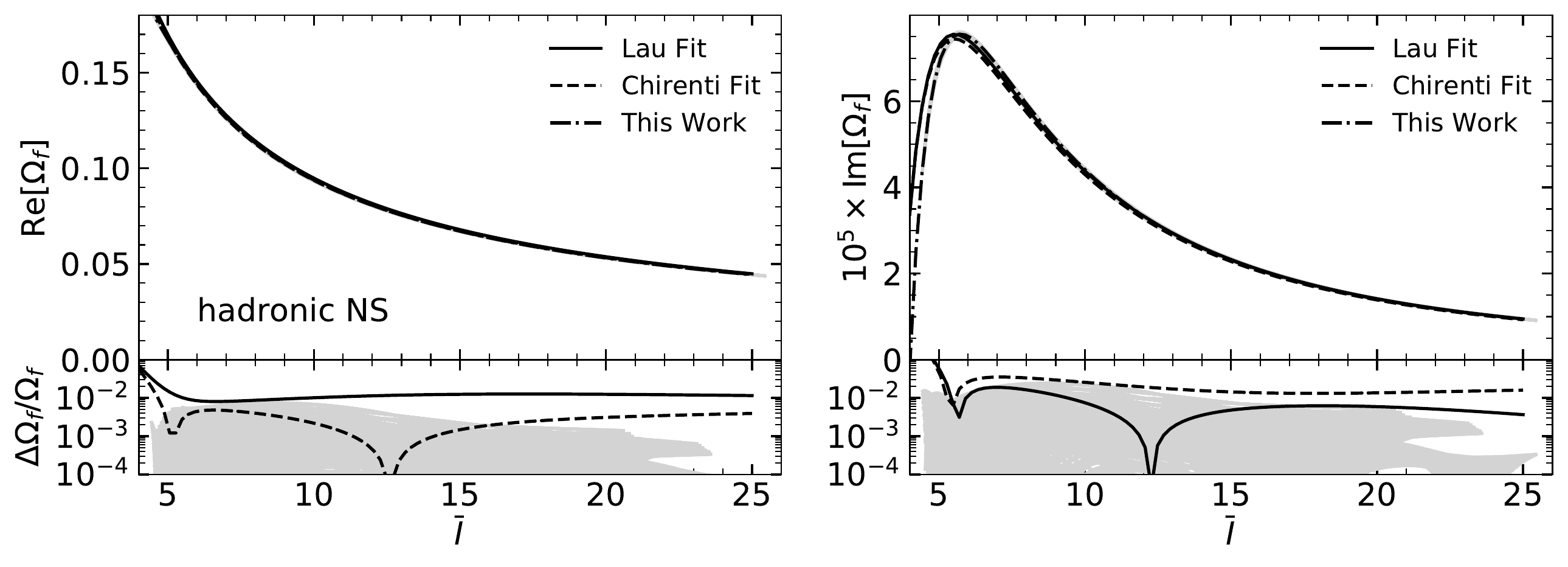}
    \includegraphics[width=\linewidth]{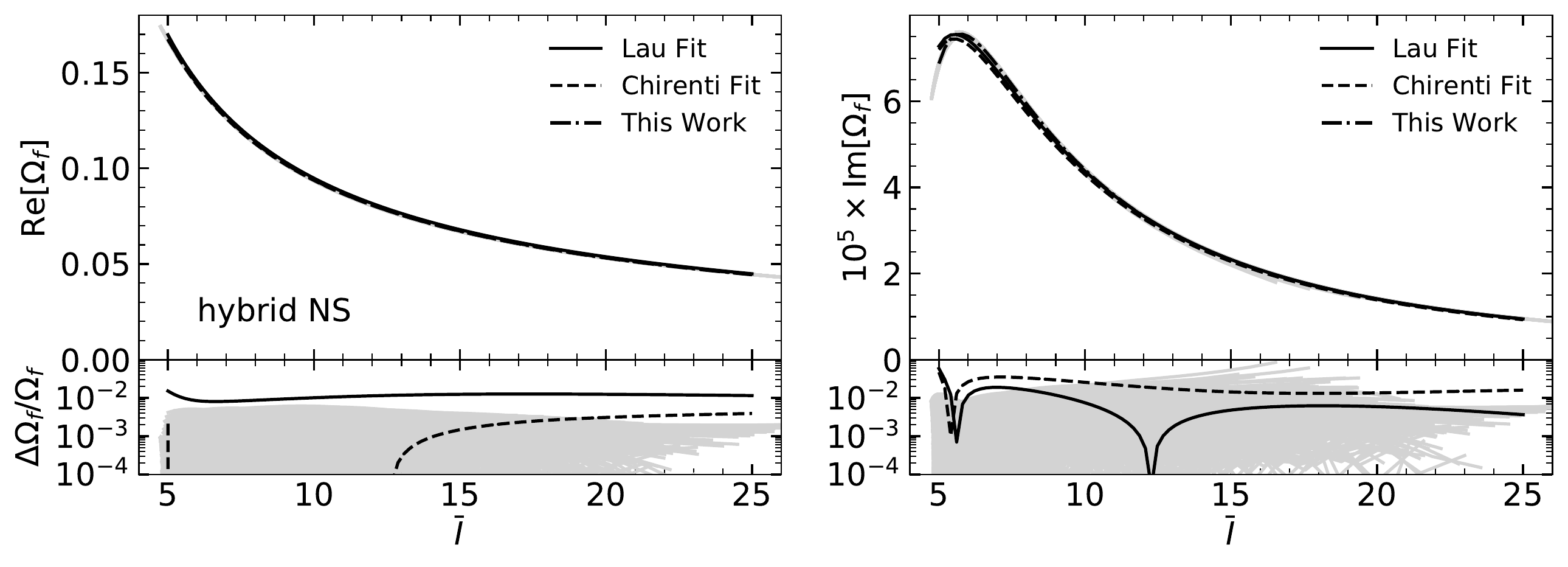}
    \includegraphics[width=\linewidth]{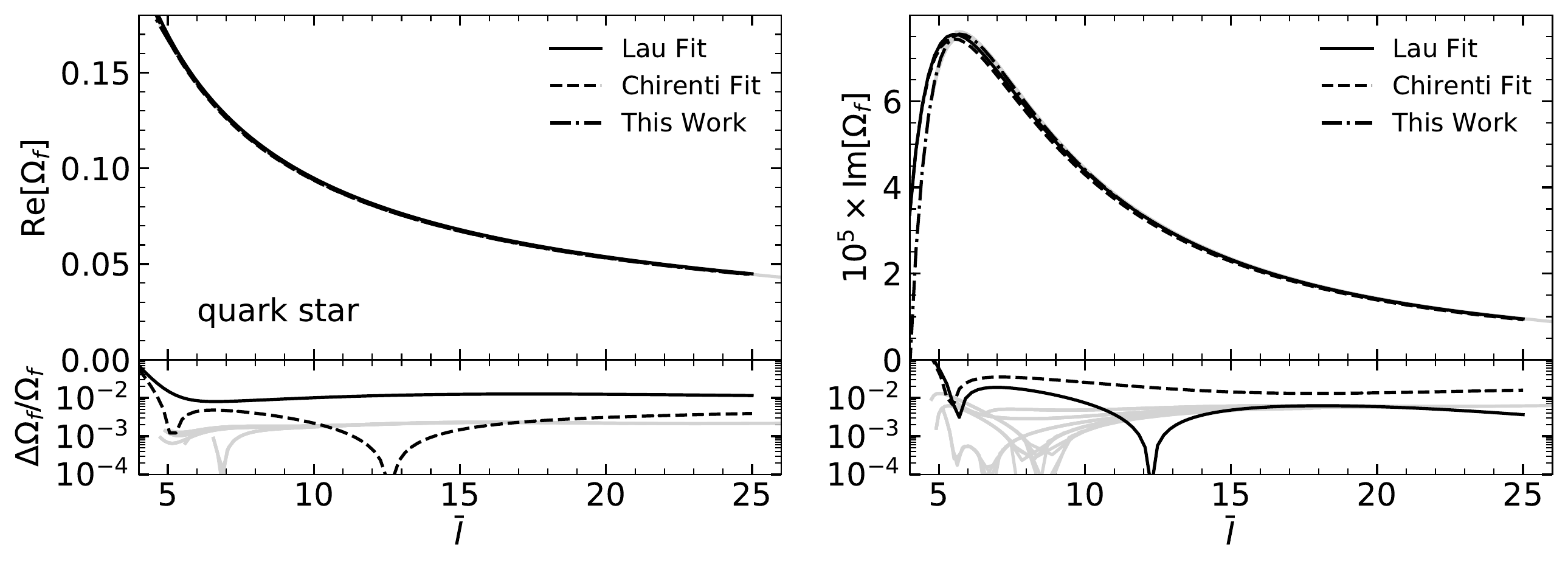}
    \caption[$f-\bar I$ EOS insensitive relation]{The same as Fig. \ref{fig:omega-beta} except for $\Omega_f-\bar I$ universal relations. ``This Work'' refers to Eq. (\ref{eq:fifit}) and other fits due to Refs. \cite{lau2010inferring} (Lau) and  \cite{chirenti2015fundamental} (Chirenti) are shown.}
    \label{fig:omega-I}
\end{figure*}
\begin{figure*}
    \centering
    \includegraphics[width=\linewidth]{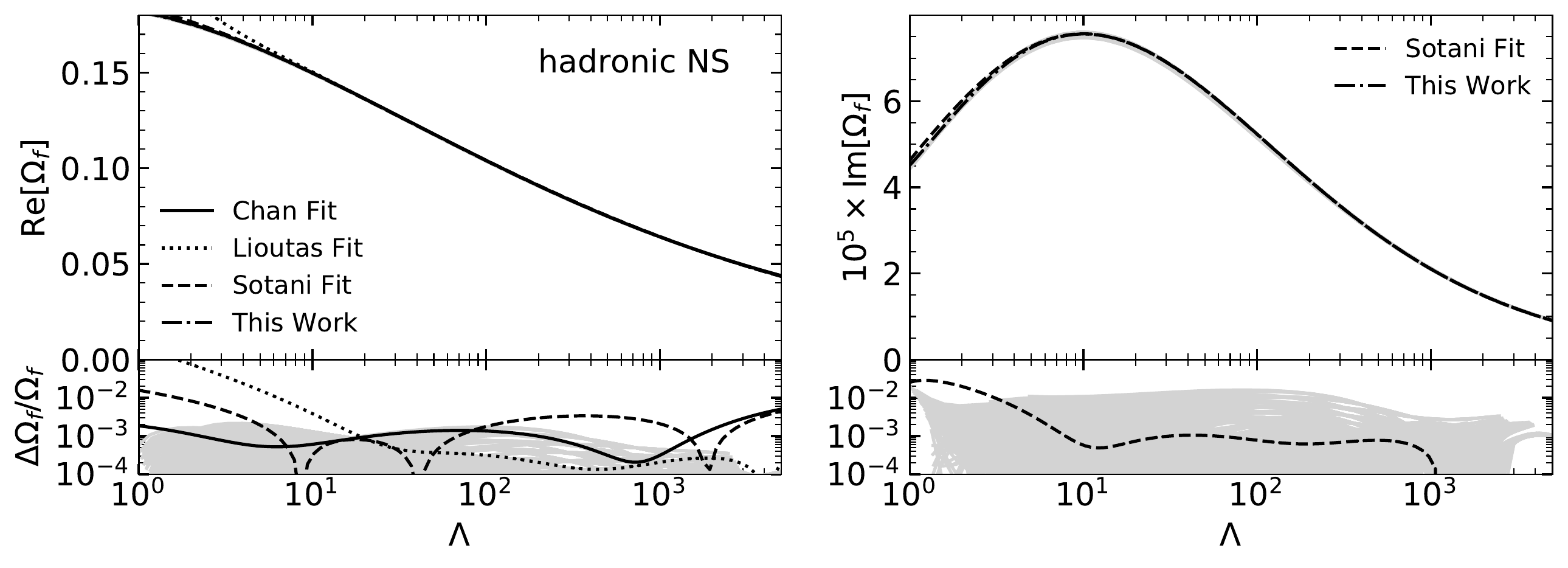}
    \includegraphics[width=\linewidth]{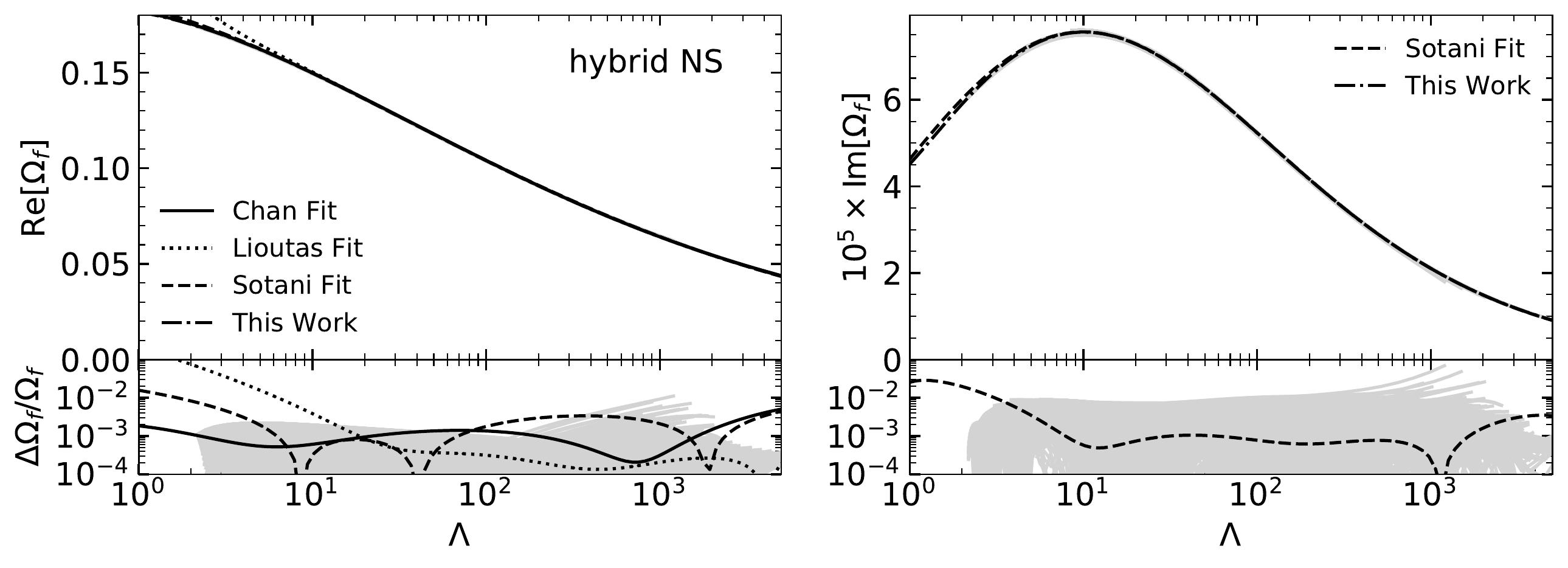}
    \includegraphics[width=\linewidth]{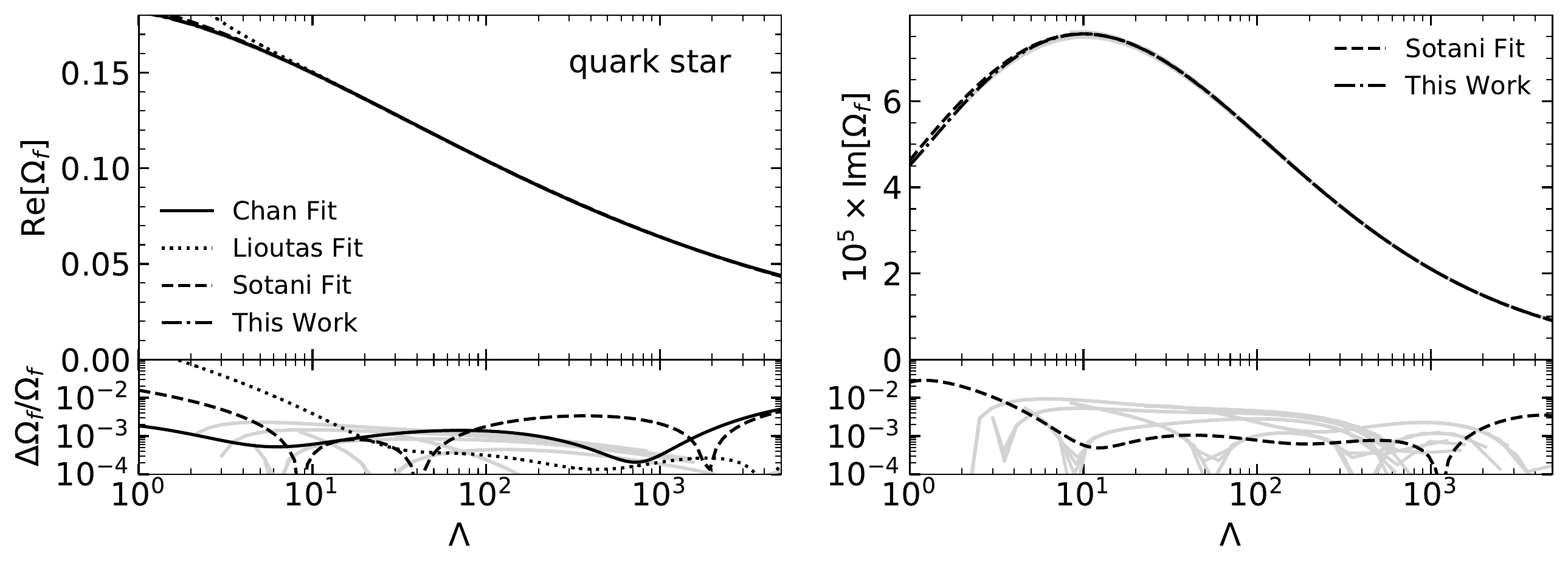}
    \caption[$f-\Lambda$ EOS insensitive relation]{The same as Fig.  \ref{fig:omega-I} but for the $\Omega_f-\Lambda$ universal relations.  ``This Work'' refers to Eq. (\ref{eq:flfit}) and other fits due to Refs. \cite{chan2014multipolar} (Chan), \cite{lioutas2021frequency} (Lioutas), and \cite{sotani2021universal} (Sotani) are shown.}
    \label{fig:omega-Lambda}
\end{figure*}

Although for Newtonian stars the relation between $\Omega_f$ and $\beta^{3/2}$ is exact, for moderate and high compactness in general realtivity it breaks down. A general power expansion of $\Omega_f(\beta)$ has shown to be relatively accurate  \cite{tsui2005universality,lioutas2021frequency}. With our parametric approach, we study the accuracy of such relations for hadronic, hybrid and quark NSs in Fig. \ref{fig:omega-beta}. We fit the hadronic and hybrid $\Omega_f-\beta$ results for $M\ge1M_\odot$ with
\begin{eqnarray}
[\Omega_{f-\beta}]_H &=& \sum_{i=0}^2 a_i \beta^i+\left(\sum_{i=4}^6 a_i \beta^i\right)\mathbf{i}
\label{eq:fbfit}\end{eqnarray}
This relation fits Re$[\Omega_f$] for hadronic stars slightly better than for hybrid NSs, and to better than 6.3\% (11.5\%) accuracy for all physically-relevant $\beta$ values. Damping times are fit to a similar precision 13.5\% for hadronic stars, but the uncertainty for hybrid stars is about 23.8\%. It is seen that the Buch and T VII solutions are good representations for both hadronic  and hybrid stars.  The $\Omega_f-\beta$ relation for frequencies and damping times for pure quark stars are much closer to the Inc analytic case, due to lack of a soft compressible crust.  For the pure quark case, we employ the fit
\begin{eqnarray}
[\Omega_{f-\beta}]_Q &=& a_0 \beta^{{3}/{2}}+\left(\sum_{i=4}^6 a_i \beta^i\right)\mathbf{i}
\label{eq:fbfitstrange}\end{eqnarray}
applicable to stars with $M\ge1M_\odot$.  The precision of the quark $\Omega_f-\beta$ relation is about 7.6\% for the real part and 6.1\% for the imaginary part. The Inc analytic model is a good representation of quark stars.  For comparison, we also show in Fig. \ref{fig:omega-beta} the fits and accuracies for hadronic stars developed by Refs. \cite{tsui2005universality} (Tsui), \cite{andersson1998towards} (Andersson), and \cite{lioutas2021frequency} (Lioutas). 

A much more precise EOS-insensitive relation relating $\Omega_f$ to $\bar I$ can be found, as first shown by  Lau et. al. \cite{lau2010inferring}, but only using a small sample of EOSs. Later, the $\bar I-\Lambda-Q$ relation was discovered \cite{yagi2013love}, so a similarly good $\Omega_f-\Lambda$ relation should also exist \cite{chan2014multipolar,chirenti2015fundamental,lioutas2021frequency}. However, Ref. \cite{wen2019gw170817} claimed quark matter EOSs violate the hadronic $\Omega_f-\bar I-\Lambda$ relation.

We tested the accuracy of the $\Omega_f-\bar I-\Lambda$ relation to the extreme for hadronic, quark and hybrid NS, and established fits valid for $M\ge0.7M_\odot$:
\begin{table*}
    \caption{Fitting parameters of real and imaginary parts of $\Omega_f$ in Eqs. (\ref{eq:fbfit}) - (\ref{eq:flfit}).\label{tab:filfit}}
    \centering
    \begin{tabular}{|c|cccccccc|}
    \hline
    &$a_0$&$a_1$&$a_2$&$a_3$&$a_4$&$a_5$&$a_6$&$a_7$\\
    \hline
    $[\Omega_{f-\beta}]_H$&-0.03044&0.6629&-0.1234&0&0.1020&-0.4752&0.5519&0\\
    $[\Omega_{f-\beta}]_Q$&0.8823&0&0&0&0.05455&-0.1912&0.1347&0\\
    Re[$\Omega_{f-\bar I}$]&0.09006&-2.41&29.47&-179.8&659.5&-1427&1689&-845.4\\
    Im[$\Omega_{f-\bar I}$]&7.506e-05&-0.002054&0.02363&-0.1484&0.5589&-1.226&1.493&-0.8139\\
    Re[$\Omega_{f-\Lambda}$]&0.1817&-0.006652&-0.004105&0.0004072&1.712e-05&-4.796e-06&2.838e-07&-5.743e-09\\
    Im[$\Omega_{f-\Lambda}$]&4.514e-05&1.907e-05&4.3e-06&-5.025e-06&1.133e-06&-1.165e-07&5.851e-09&-1.167e-10\\
    \hline
    \end{tabular}
\end{table*}
\begin{eqnarray}
\Omega_{f-\bar I} &=& \sum_{i=0}^6 a_i \bar I^{-{i}/{2}}\label{eq:fifit}\\
\Omega_{f-\Lambda} &=& \sum_{i=0}^6 a_i (\ln{\Lambda})^i
\label{eq:flfit}\end{eqnarray}
The fitting parameters are given in Table \ref{tab:filfit}.

Figs. \ref{fig:omega-I} and \ref{fig:omega-Lambda} show the $\Omega_f-\bar I-\Lambda$ relations for hadronic, hybrid and quark NS. No hybrid nor quark star significantly violates the $\Omega_f-\bar I-\Lambda$ fits in Eqs. (\ref{eq:fifit}) and (\ref{eq:flfit}) with a few exceptions of low-mass hybrid stars $\lesssim1M_\odot$ with strong phase transitions.   Otherwise, the accuracy is better than about 1\%. The correlation for the real part is somewhat more accurate than for the imaginary part. For $\bar I\gtrsim 15$, i.e., $M \lesssim 1.6 M_\odot$, the maximum deviation of the real part reduces to only 0.3\%. Interestingly, the $\Omega_f-\Lambda$ relation is generally even more precise than for the $\Omega_f-\bar I$ relation except for low-mass configurations near $1M_\odot$. In hadronic NSs, the maximum deviation of the real (imaginary) part is about 0.2\% (2\%). 
The fits due to Refs. \cite{lau2010inferring} (Lau), \cite{chirenti2015fundamental} (Chirenti),  \cite{chan2014multipolar} (Chan), \cite{lioutas2021frequency} (Lioutas), and \cite{sotani2021universal} (Sotani) are also shown.

\section{The special 1-node f-mode branch\label{sec:nodes}}
\begin{figure*}
    \centering
    \includegraphics[width=0.43\linewidth]{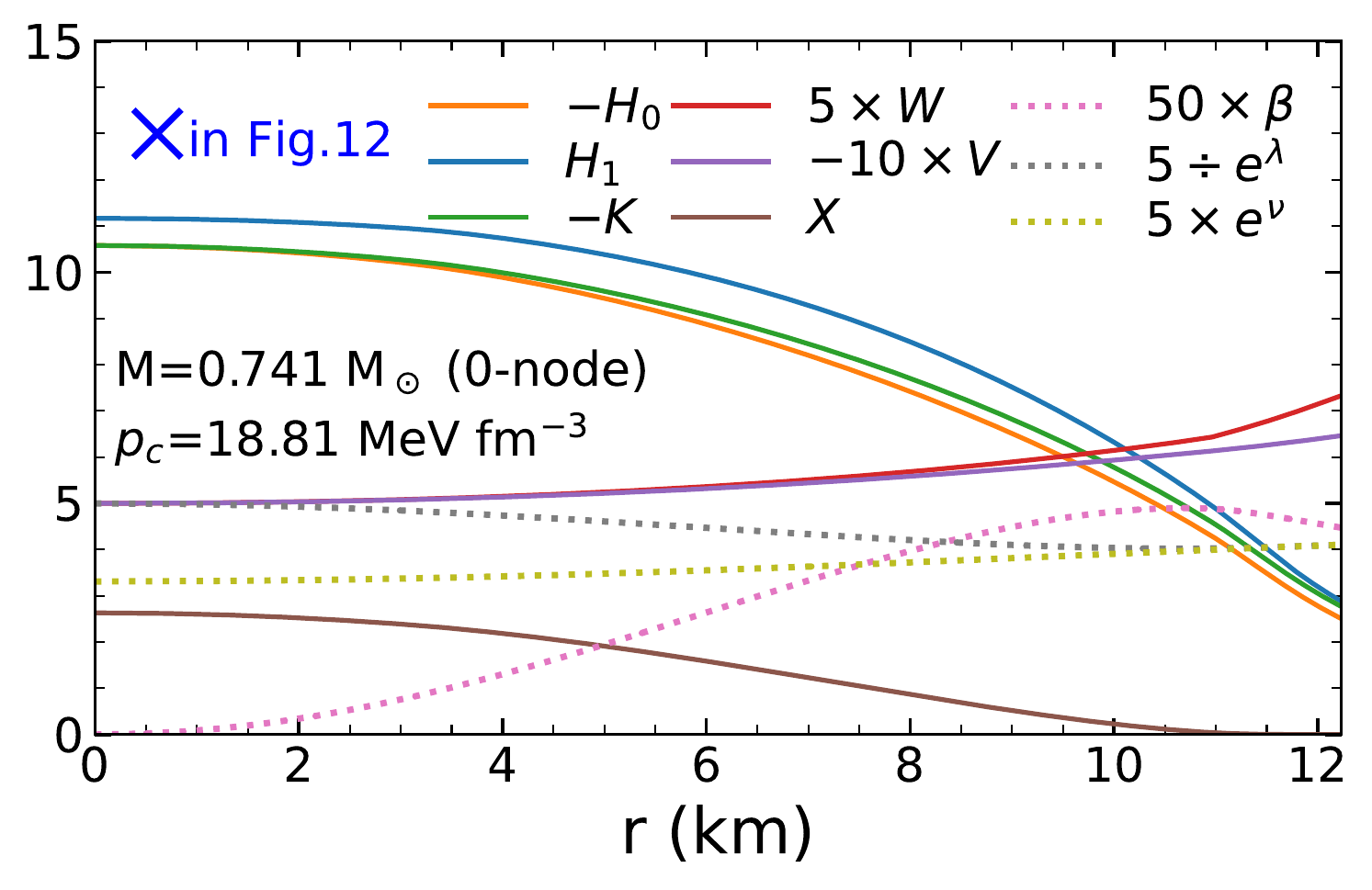}
    \includegraphics[width=0.43\linewidth]{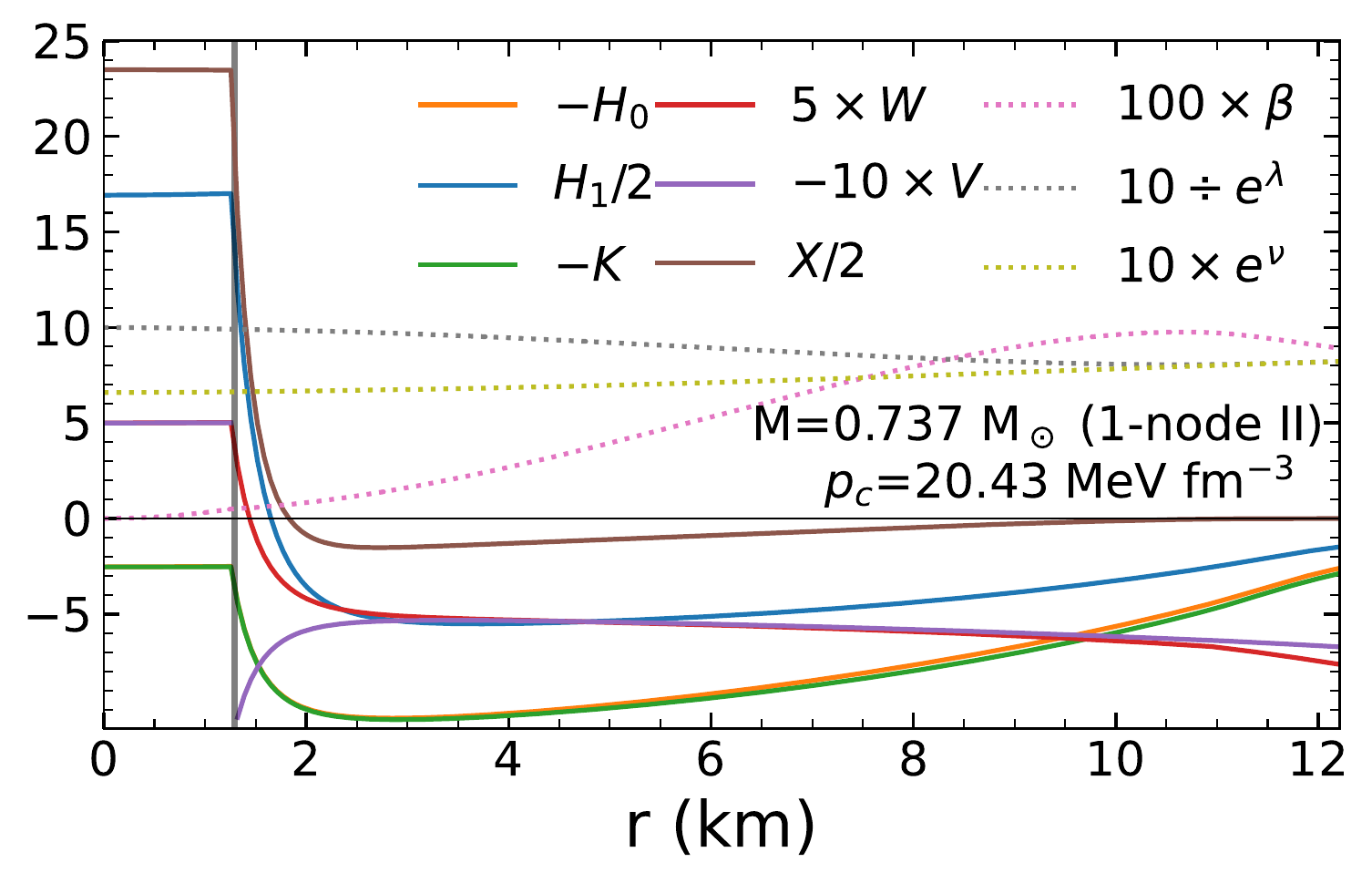}
    \includegraphics[width=0.43\linewidth]{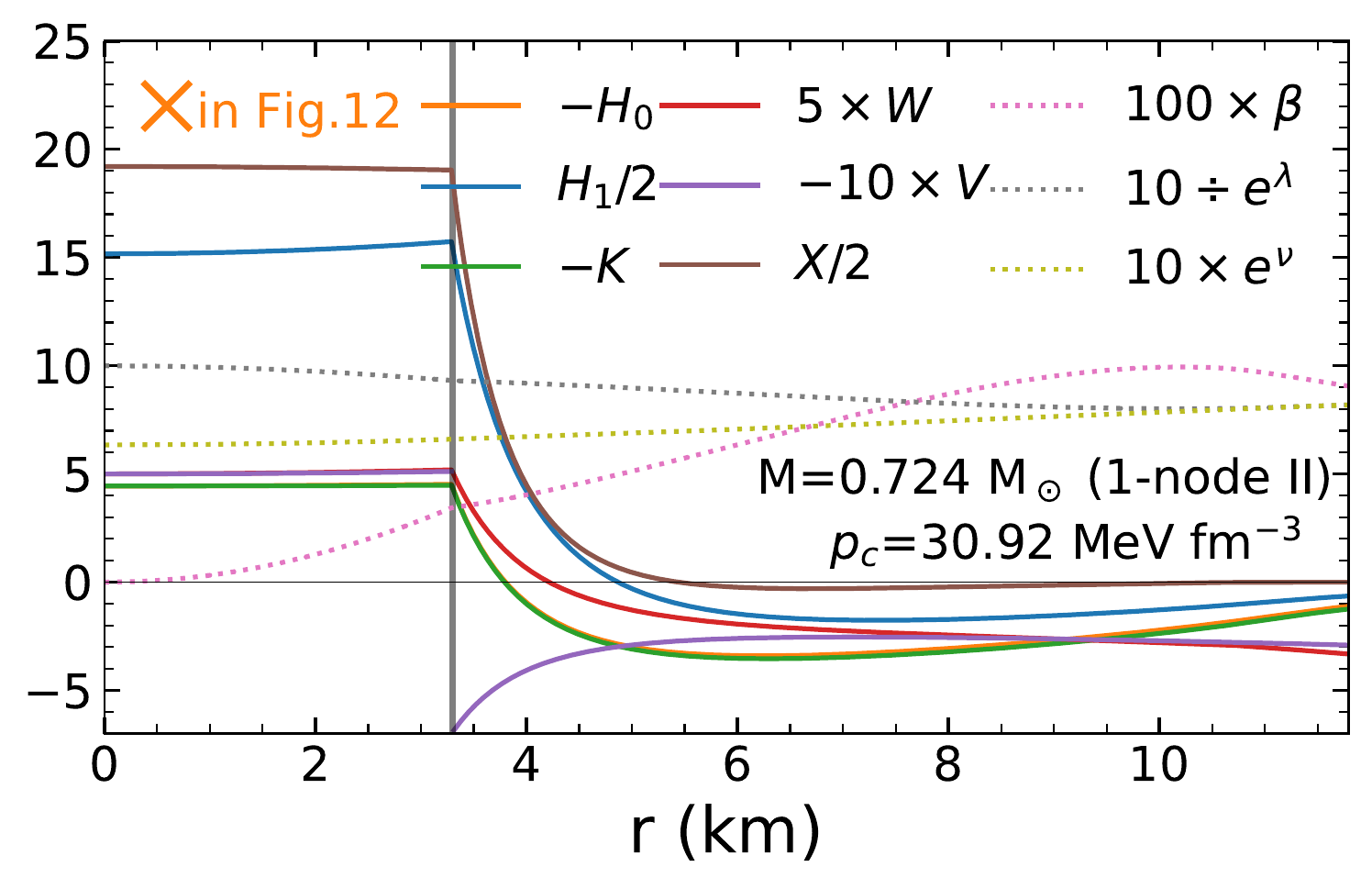}
    \includegraphics[width=0.43\linewidth]{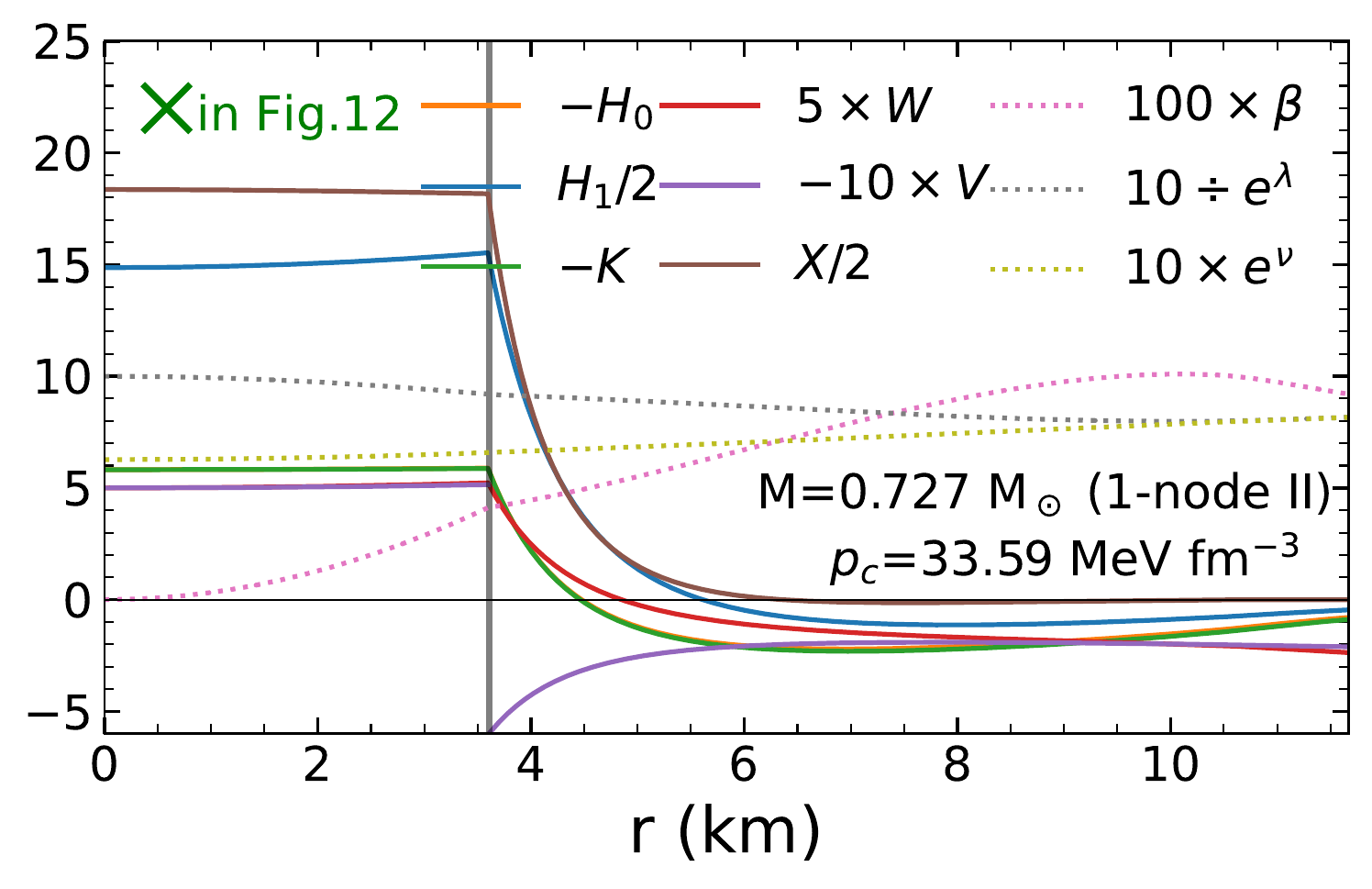}
    \includegraphics[width=0.43\linewidth]{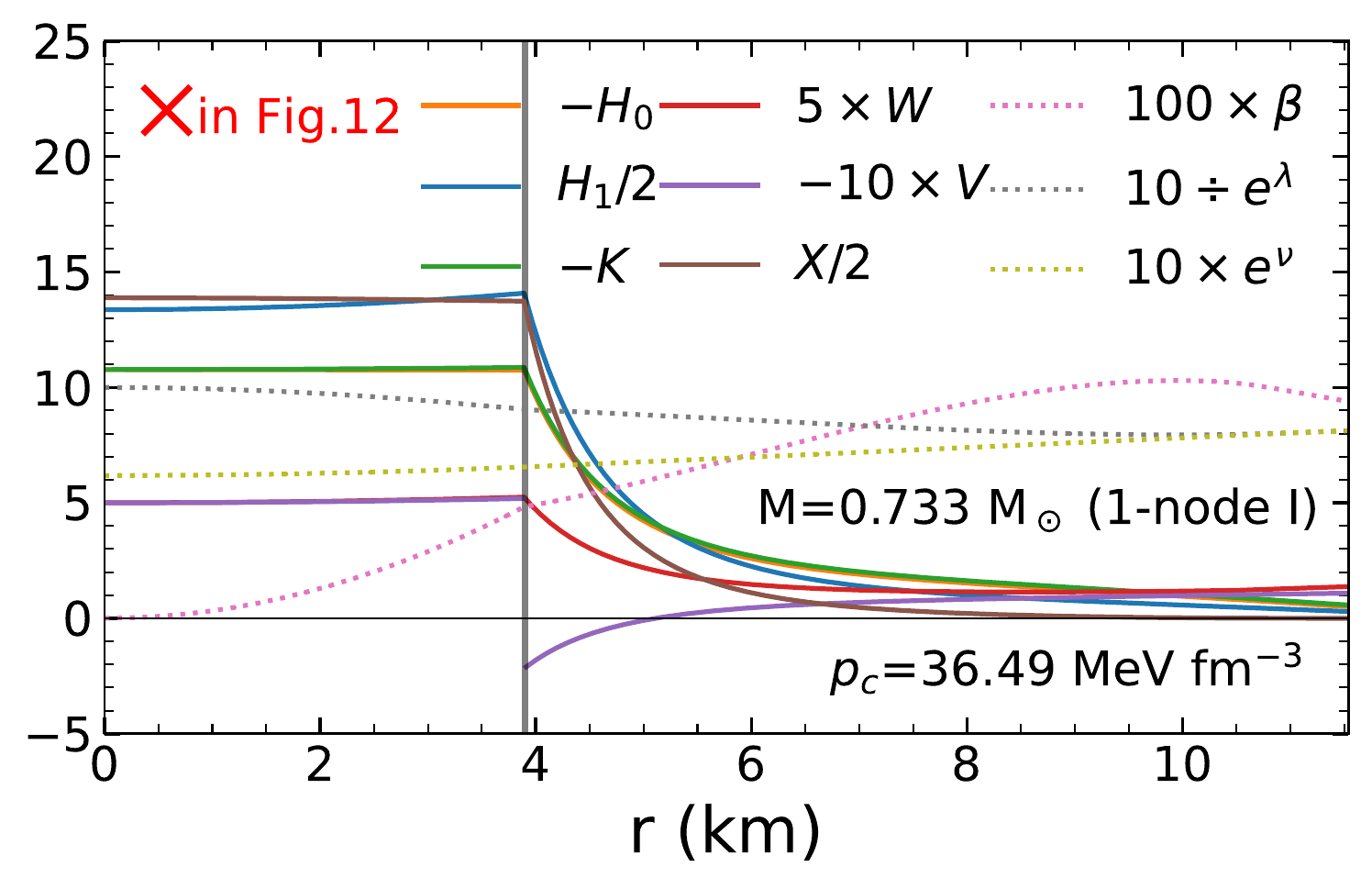}
    \includegraphics[width=0.43\linewidth]{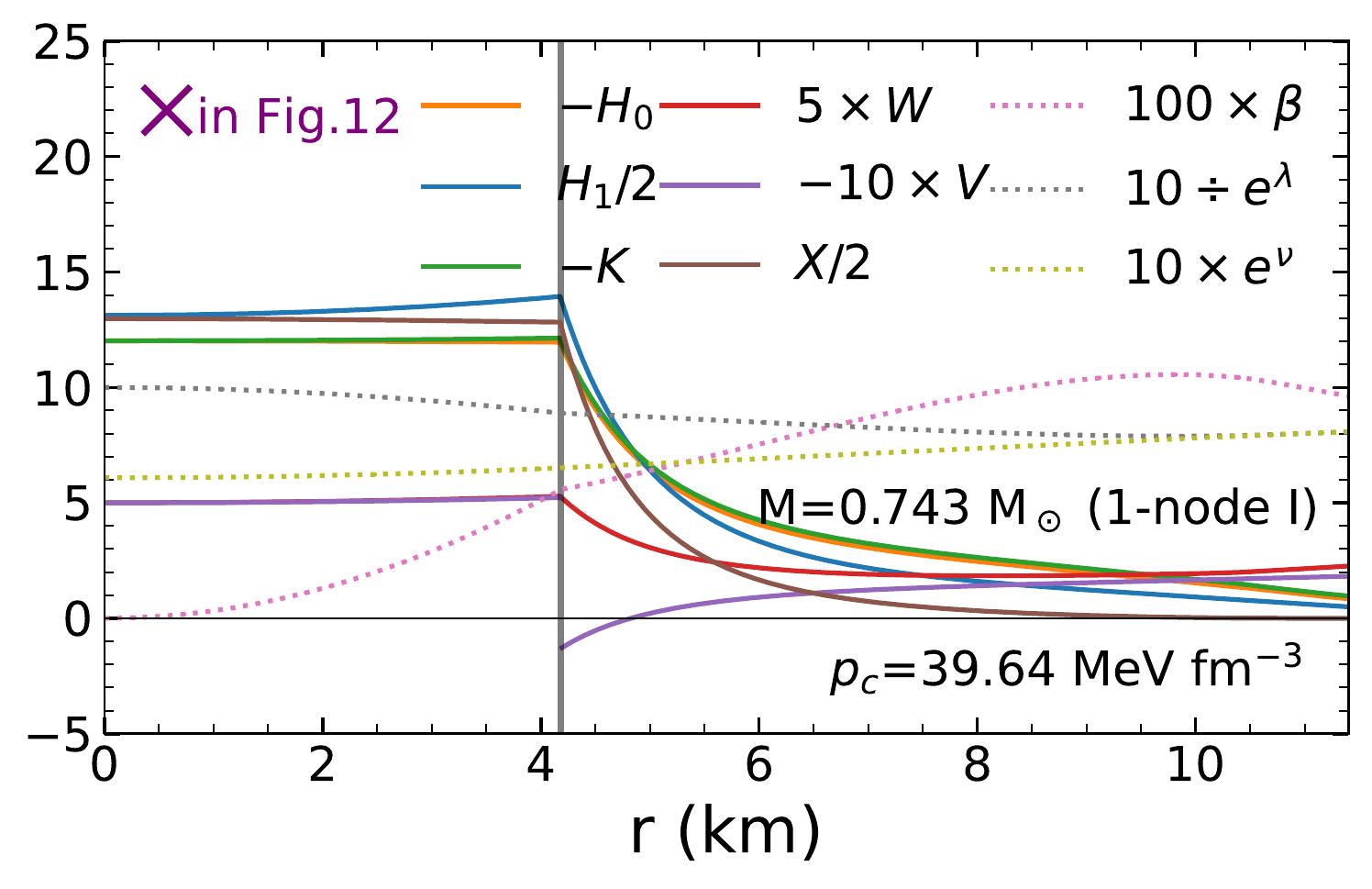}
    \includegraphics[width=0.43\linewidth]{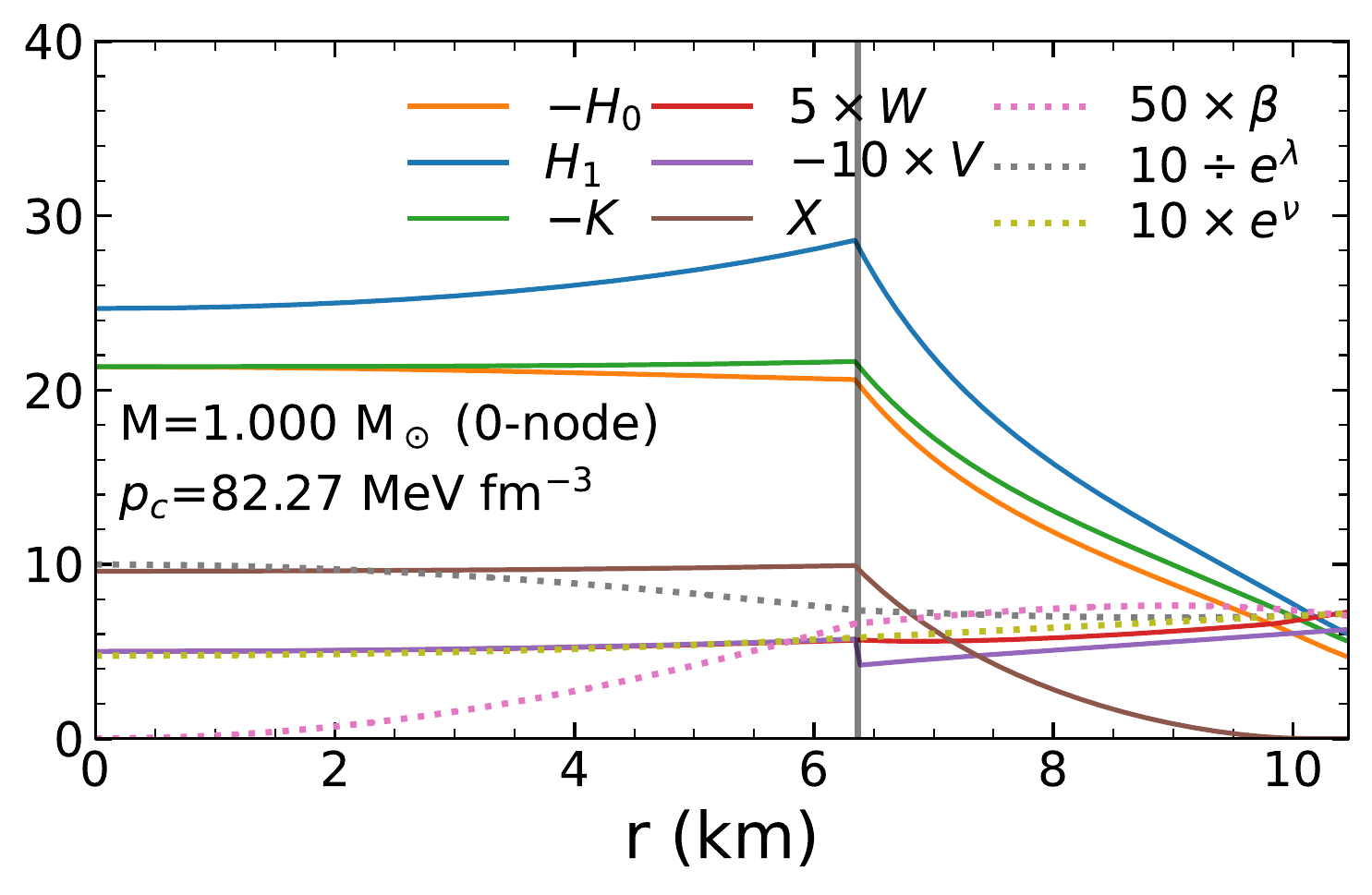}
    \includegraphics[width=0.43\linewidth]{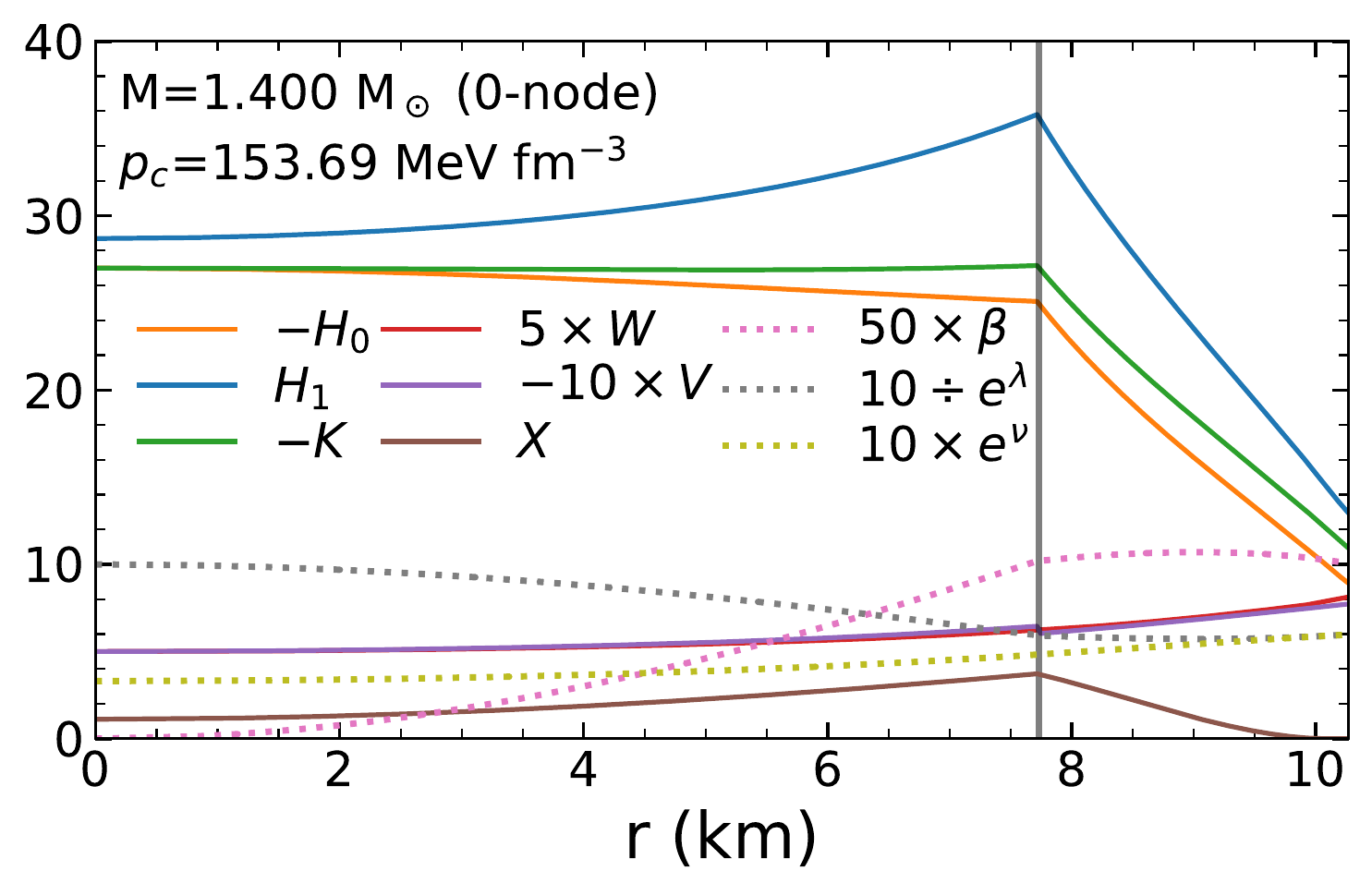}
    \caption[Metric perturbation functions and fluid displacement functions when the star contains a first-order transition]{Metric and real parts of fluid perturbation amplitudes (solid), and static metric functions (dotted) for the case of a hybrid star (except for the upper left panel where the central pressure equals the transition pressure, $p_c=p_t$). The phase transition is marked with a vertical solid line at $r=R_t$. Five of the panels have the same $p_c$ as in Fig. \ref{fig:f-mode_two_branches} and are so indicated with colored crosses.  $H_0$, $H_1$ and $K$ are in units of $\varepsilon_s=152.26$ MeV fm$^{-3}$, $X$ is in units of $\varepsilon_s^2$ and $W,V,\lambda$ and $\nu$ are dimensionless.   \label{fig:HKWX_hybrid_branches}}
\end{figure*}
\begin{figure*}
    \centering
    \includegraphics[width=\linewidth]{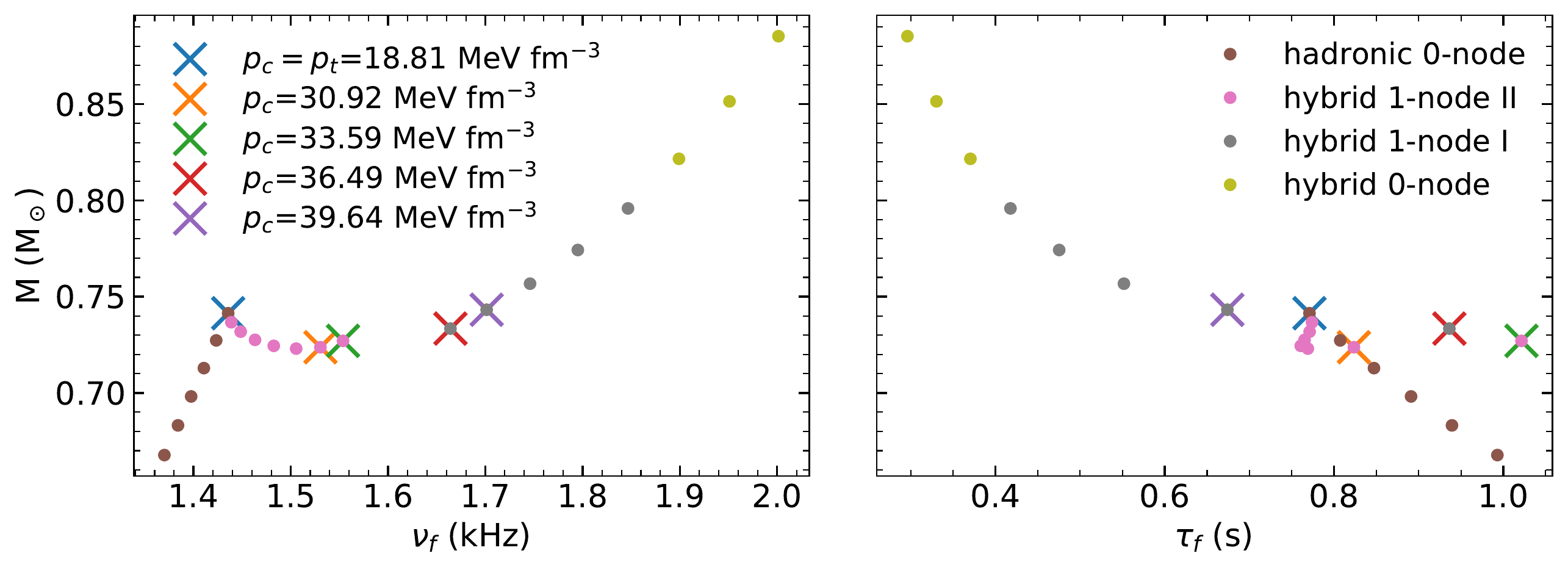}
    \includegraphics[width=\linewidth]{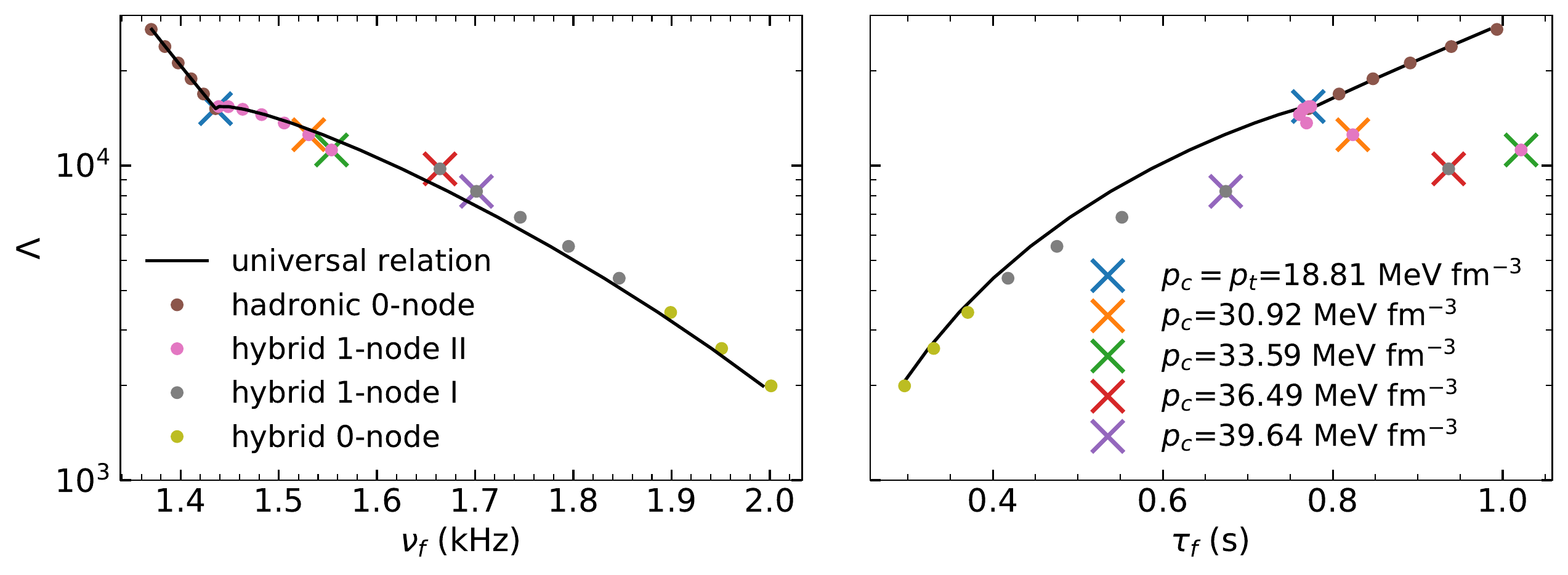}
    \caption{f-mode frequency and damping time for hybrid NSs with small quark cores as functions of mass and tidal deformability. Brown and yellow dots represent stars with normal f-modes having no radial nodes, while pink and grey dots represent stars with the special 1-mode behavior. The colored crosses correspond to 5 configurations displayed in Fig. \ref{fig:HKWX_hybrid_branches}. In the tidal deformability plots, the universal relations from Eq. (\ref{eq:flfit}) are shown. \label{fig:f-mode_two_branches}}
\end{figure*}

The f-mode is characterized by the lowest radial order (n=0) for normal EOS. We found the f-modes of low-mass hybrid stars with strong phase transition that lead to the so-called twin star phenomenon can be qualitatively different than the normal f-modes.  The twin star phenomenon is the situation where, as central density is increased through the phase transition density, the mass and radii both initially decrease. Such configurations are dynamically unstable.  As the central density is further increased, the mass may begin to rise while the radius continues to fall, leading to a twin branch of stable configurations.  Note that the normal and twin stable branches are disconnected in $M-R$ space.  It is then possible to have stable stars with the same mass but differing radii (and also f-mode frequencies).  Such a situation can be seen in Fig. \ref{fig:f-mode_hybrid} for a few cases, including $c_s^2=1, n_t=2n_s$ and $\Delta\varepsilon/\varepsilon_t=0.8$ (solid red curves in lowest panels).  To demonstrate the different f-mode behavior on the more compact twin star branch, we utilize this particular EOS in Figs. \ref{fig:HKWX_hybrid_branches} and \ref {fig:f-mode_two_branches}.

In a normal star without a first-order transition, all perturbation amplitudes are continuous and smooth, see Fig. \ref{fig:HKWX_EOS_SLY4_1.4} or the top left panel of  Fig. \ref{fig:HKWX_hybrid_branches}. 
In a hybrid star with a first-order transition, however, the slopes of all perturbation amplitudes become discontinuous, and $V$ itself becomes discontinuous, at the transition boundary  (last 7 panels of Fig. \ref{fig:HKWX_hybrid_branches}). For hybrid stars with relatively small quark cores that occupy the twin star branch, all fluid and metric perturbation amplitudes can become negative in some parts of the star between the phase transition and the surface (second through sixth panels of Fig. \ref{fig:HKWX_hybrid_branches}. The overall sign of the perturbations is trivial since we define a positive fluid perturbation amplitude $W=1$ at the center of the NS. What's nontrivial is that the unstable and some stable hybrid stars have a radial node (zero) in the fluid and metric perturbation amplitudes $X$, $W$, $H_0$, $H_1$, $K$ (but not $V$, which, however discontinuously changes sign) at a radius slightly larger than the phase transition radius $R_t$. We will call this type of behavior 1-node II (second through fourth panels of Fig. \ref{fig:HKWX_hybrid_branches}).  The radial nodes move outward with increasing central density or pressure and $R_t$. Above some critical $R_t$, the radial nodes for $X$, $W$, $H_0$, $H_1$ and $K$ simultaneously vanish, but the discontinuity and sign change in $V$ remains (fifth and sixth panels of Fig. \ref{fig:HKWX_hybrid_branches}). Stars with radial nodes in $V$ only we refer to as 1-node I. For hybrid stars with even larger cores, the discontinuity in $V$ remains, but it no longer has a sign change at $R_t$, and the f-mode oscillations recover the standard $n=0$ behavior of hadronic stars (last two panels of Fig. \ref{fig:HKWX_hybrid_branches}). Although $V$ is always discontinuous at $R_t$, see Eq. (\ref{eq:V_def_in}), the discontinuity magnitude becomes small for massive hybrid stars. 
The exotic behavior of the perturbation amplitudes causes f-mode frequencies of twin stars to somewhat deviate from universal relations for masses near $M_t$, see Fig. \ref{fig:f-mode_two_branches}. A gap of about 0.1 kHz can appear in the f-mode frequency spectrum for $M\simeq M_t$. The f-mode frequencies and damping times of both 0-node and 1-node hybrid stars show deviations as well, see middle panels of Fig. \ref{fig:omega-Lambda} and Fig. \ref{fig:omega-I}. The f-mode frequency of hybrid NSs is not continuous between the 1-node I and 1-node II branches. The 1-node II branch side has a lower frequency than the 1-node I branch, and the damping time greatly increases near the critical point between 1-node I and 1-node II branches.
 
While the particular configurations explored here would never be expected to be observed because of the small value of $M_t$, such behavior could have observational consequences for hybrid EOSs where $M_t\gtrsim1M_\odot$.  For example, the case $c_s^2=1, n_t=4n_s$ and $\Delta\varepsilon/\varepsilon_t=0.8$ shown in Fig. \ref{fig:f-mode_hybrid} (red dashed line in the bottom row) has a transition mass $M_t\simeq1.6M_\odot$.  However, even this case cannot be observationally realized since $M_{\rm max}\simeq1.6M_\odot$ is too small.  Indeed, for the N3LO-cen and $\pm\sigma$ hadronic EOSs, which are relatively soft, we do not find it possible to produce twin stars simultaneously having $M\ge1M_\odot$ and $M_{\rm max}\ge2M_\odot$. It is possible, however, for these conditions to be realized for stiffer hadronic EOSs.  The models of Refs.  \cite{montana2019constraining} and \cite{Fattoyev2020}  provide some examples for which the hadronic EOSs can both satisfy saturation density symmetry energy and GW170817 tidal deformability constraints. One caveat is that the quark matter EOSs must have $c_s^2>1/3$. The twin star cases in these studies are of two types, one having $M_t$  below that of the largest component of GW170817 but $\ge1M_\odot$, and the other having $M_t\gtrsim 2 M_\odot$. Therefore the 1-node behavior we find might be potentially observable.  In any case, our previously established universal relations can be accurately extended to hybrid stars, with small violations in the case of twin stars with masses near $M_t$.
\section{The discontinuous g-mode of Hybrid NS\label{sec:g-mode}}
\begin{figure*}
    \centering
    \includegraphics[width=\linewidth]{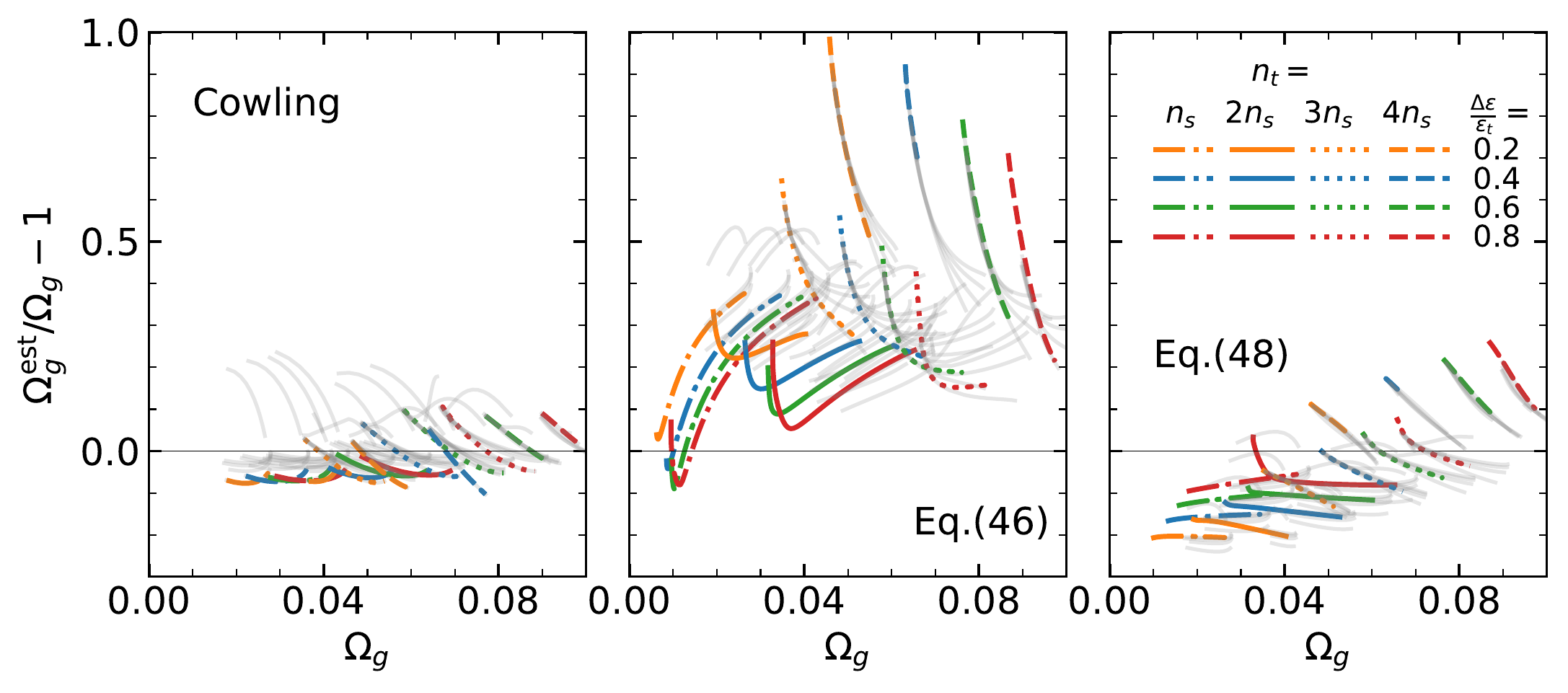}
    \caption[Comparison between approximated (y-axis) and numerically calculated (x-axis) g-mode frequency under full GR]{Deviations of various estimations of g-mode frequencies from fully general relativistic calculations. The left, middle and right panels show the Cowling approximation and the fits of Eq. (\ref{eq:g_mode_esti_finn})  and (\ref{eq:g_mode_esti_my}), respectively.  Colored lines correspond to cases with $c_s^2=1/3$ while gray lines show results with higher values of $c_s^2$.} \label{fig:g_mode_estimate}
\end{figure*}

We are also interested in the g-mode frequency permitted by a discontinuity inside a NS. Since the EOSs of both strange quark stars and hadronic stars have no discontinuities, their g-mode frequencies are trivially zero. In this section, we focus on the discontinuous g-mode for hybrid stars described in Section \ref{sec:EOS}. We calculated g-mode frequencies with and without the Cowling approximation (left panel of Fig. \ref{fig:g_mode_estimate}). The relative error reaches 12\% (22\%) for $c_s^2=1/3$ ($c_s^2\ge1/3$). This error is significantly lower than that of the Cowling approximation for the f-mode. Considering that there's no universal relation accurate to the few percent level for the g-mode, the Cowling approximation can be a reasonable approximation. However, previous studies have been too optimistic about its accuracy, e.g. a claimed 5\% error by Refs. \cite{sotani2001density} and \cite{finn1987g}. Partly, this is due to our consideration of a wider variety of first order transitions, not just the relatively weak core-crust transition, as well as our inclusion of more realistic NSs ($M>1.2 M_\odot$).  

Furthermore, given that even the Cowling approximation involves a complex numerical calculation, it is useful to find an analytic fit to the general relativistic results.  In Newtonian fluid mechanics, the frequency of surface gravity waves between two stratified fluids with a uniform gravitational field (i.e., the slab approximation) is analytically solvable \cite{landau2013fluid}:
\begin{eqnarray}
\omega^2_g= \frac{(\varepsilon_+-\varepsilon_-)gk}{{\varepsilon_+}/{\tanh{[kd_+]}}+{\varepsilon_-}/{\tanh{[kd_-]}}}, \label{eq:g_mode_two_layer}
\end{eqnarray}
where $g$ is the gravitational acceleration and k is the angular wave number. $\varepsilon_+=\varepsilon_t+\Delta\varepsilon$ ($\varepsilon_-=\varepsilon_t$) and $d_+=R_t$ ($d_-=R-R_t$) stand for the energy density and depth, respectively, on the high (low) density side. When the discontinuity at $R_t$ happens near the surface $R$ of a star, the geometry approximates a stratified two-fluid problem with $k=\sqrt{l(l+1)}/R_t$.  Red shifts in the frequency and gravitational acceleration approximately cancel, so $g$ can be taken to be that of Newtonian gravity, $GM_t/R_t^2$.  Ref. \cite{finn1987g} concludes  that discontinuous g-modes near the NS surface can be approximated, using $R-R_t<<R_t$, $M-M_t<<M_t$, and $\Delta\varepsilon<<\varepsilon_t$, as
\begin{eqnarray}
\Omega^2_g &\approx& l(l+1)\beta^3 \frac{\Delta\varepsilon/\varepsilon_t}{1+\Delta\varepsilon/\varepsilon_t} (1-R_t/R). \label{eq:g_mode_esti_finn}
\end{eqnarray}
  We tested this approximation with $\ell=2$ and find relatively large deviations, see the middle panel of Fig. \ref{fig:g_mode_estimate}, partly because their assumption of infinite depth $d_+>>d_-$ breaks down for $R_t/R<0.5$ \cite{miniutti2003non}.  Instead, 
we don't assume $\Delta\varepsilon<<\varepsilon_t$, $M-M_t<<M_t$ or $R-R_t<<R_t$, and we also approximate the wave number with $k=D/R_t$, where the fitting parameter $D=1.21$.  This leads to
\begin{equation}
\Omega^2_g\approx \frac{\beta^3 (M_t/M)(R/R_t)^3(\Delta\varepsilon/\varepsilon_t)D\tanh[D]}{1+\Delta\varepsilon/\varepsilon_t+ \tanh[D]/\tanh[D(R/R_t-1)]}, \label{eq:g_mode_esti_my}
\end{equation}
which performs significantly better than Eq. (\ref{eq:g_mode_esti_finn}) and comparably to the Cowling approximation, see Fig. \ref{fig:g_mode_estimate}.  
We note these fits for the g-mode frequency depends only on $M/R,M_t/M,R_t/R$ and $\Delta\varepsilon/\varepsilon_t$ and are otherwise insensitive to the hybrid EOS parameters $\varepsilon_t$ and $c_s^2$ as well as the assumed hadronic EOS.

\begin{figure*}
    \centering
    \includegraphics[width=\linewidth]{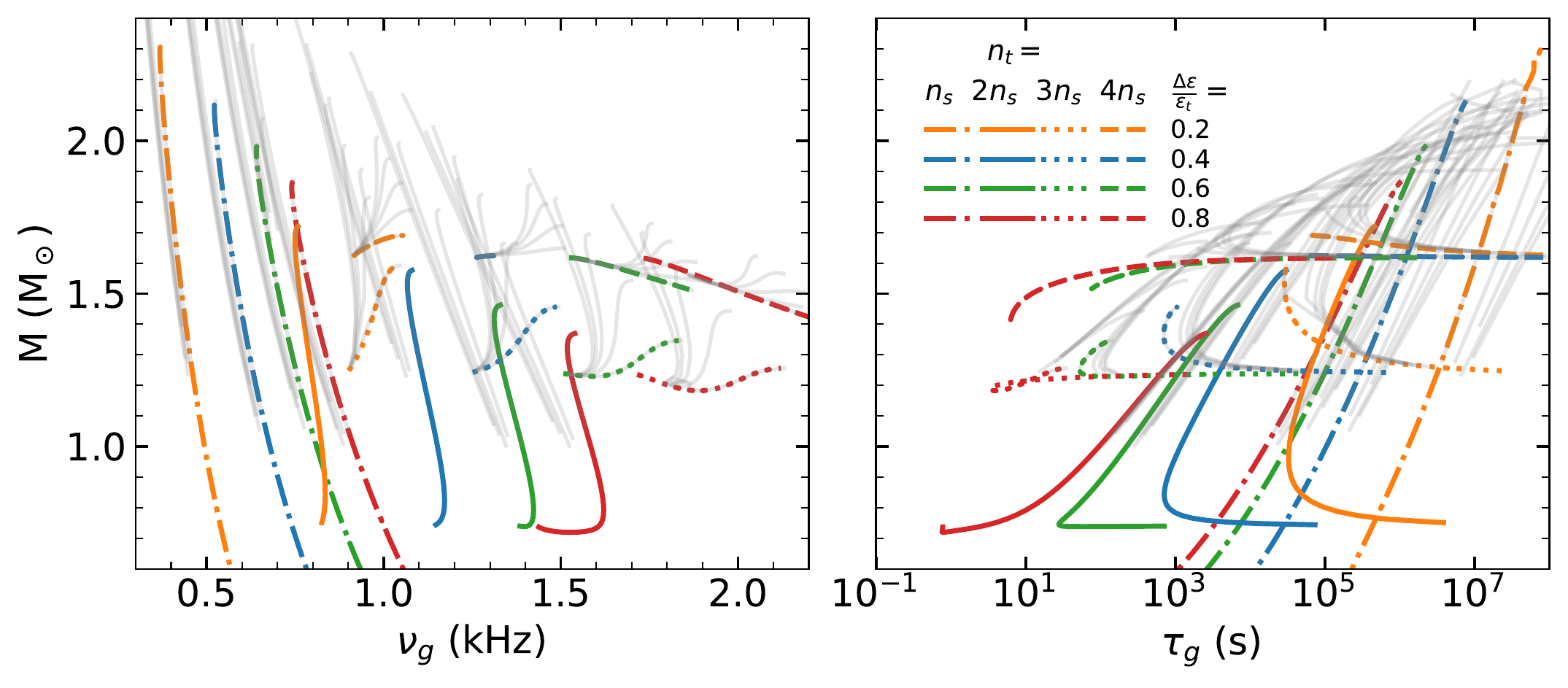}
    \caption[g-mode frequency and damping time versus mass with $\chi$EFT and \emph{CSS}]{The left (right) panel shows the g-mode frequency (damping time) versus mass for hybrid NSs modeled with \chiEFT and the \emph{CSS} parameterization. Colored lines correspond to cases with $c_s^2=1/3$ while gray lines show results with higher values of $c_s^2$.  \label{fig:g_mode_transition}}
\end{figure*}
Fig. \ref{fig:g_mode_transition} shows the mass dependence of the g-mode frequency and damping time with various high-density sound speeds, transition densities and density discontinuities. In most cases for stable NSs, g-mode frequencies are not very sensitive to NS mass. However, damping times have a very strong mass dependence. Both very low and very high mass hybrid NSs have relatively long damping times.  A previous study with a different EOS parameterization suggested that g-mode damping times  are significantly larger than those of other damping mechanisms  \cite{miniutti2003non}. Our calculation shows smaller g-mode damping times. When the density discontinuity approaches $\Delta\varepsilon/\varepsilon_t=1$, g-mode damping times become comparable to the neutrino damping time, $0.1-10$ s  \cite{alford2019damping}. However, these configurations have parameterizations with relatively low $M_{\rm max}$. If we impose $M_{\rm max}>2M_\odot$ and causality constraints, the discontinuous g-mode damping times should satisfy $\tau_g\gtrsim10^6$ s.

\begin{figure*}
    \centering
    \includegraphics[width=\linewidth]{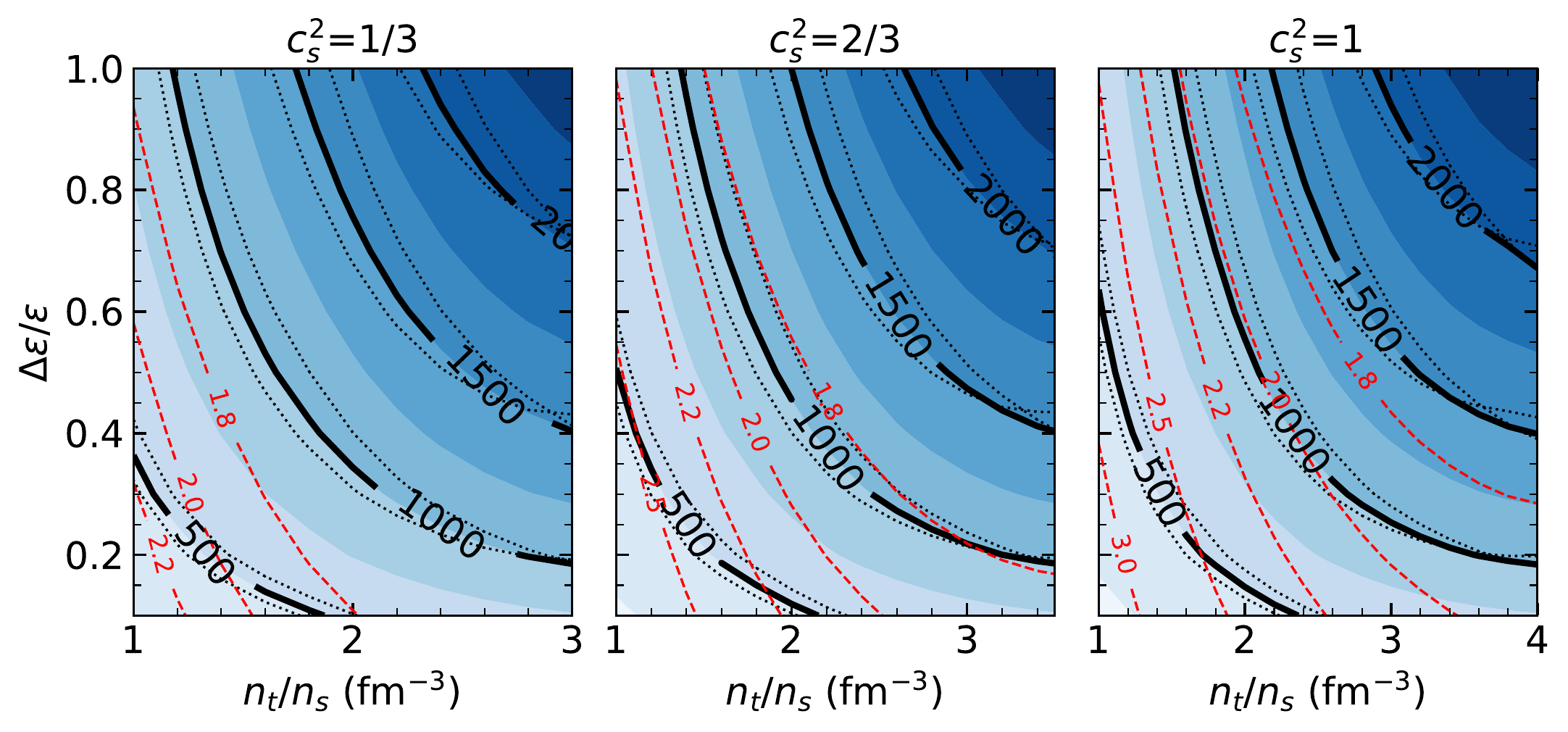}
    \caption[Parameter space of \emph{CSS} with contour of mass and g-mode frequency of the maximal massive NS.]{The g-mode frequencies $\nu_{g,\rm max}$ of maximum mass hybrid configurations are shown as black contours using the hadronic N3LO-cen chiral EFT EOS, with dotted lines representing the $\pm\sigma$ uncertainties in the hadronic EOS.  Intermediate frequency values are indicated by blue shading.  Red contours indicate $M_{\rm max}$.}
    \label{fig:g_mode_frequency}
\end{figure*}
Since the g-mode frequency is insensitive to NS mass (Fig. \ref{fig:g_mode_transition}), we focus on the maximum mass configurations for hybrid NS EOSs, shown in Fig. \ref{fig:g_mode_frequency}. If we additionally impose the M$_{max}>2M_\odot$ and causality constraints, the discontinuous g-mode frequency should satisfy $\nu_g\lesssim1.25$ kHz. A fitting formula for the g-mode frequency of hybrid $M_{\rm max}$ NSs is
\begin{eqnarray}
\nu_{g,\rm max} &=& a_g \left(\frac{p_t}{\textrm{MeV fm}^{-3}}\right)^{\gamma_g} \sqrt{\frac{\Delta\varepsilon}{\varepsilon_t}} \left(\frac{c}{c_s}\right).
\end{eqnarray}
where $a_g=326.4\pm 36.1 \textrm{Hz}$ and $\gamma_g=0.2683+0.1462 {c_s}/{c}$. g-modes of observed stars will be at larger frequencies.
\section{Discussion and Conclusion}
With the improvement provided by the next generation of gravitational-wave telescopes, we may detect gravitational waves from quasi-normal modes of NS oscillation. In this paper, we focus on the f-mode and the discontinuity g-mode which frequencies in the detectable range. The f-mode has a relatively large coupling with tidal excitations, while the discontinuity g-mode has a lower frequency that would be excited earlier in the inspiral stage of a BNS merger and which would give a larger phase shift due to the additional orbital momentum decay. 

The NS oscillations were calculated using three methods. In asteroseismology, fluid oscillations, including g-, f- and p-modes, have been extensively studied in Newtonian gravity \cite{cox2017theory}. In compact object where general relativistic effects are important, the canonical ODEs for non-radial oscillation were proposed by \cite{thorne1967non} and reformulated by Refs. \cite{lindblom1983quadrupole,detweiler1985nonradial} in full general relativity and by Ref.  \cite{mcdermott1983nonradial} utilizing the relativistic Cowling approximation.

We first compared Newtonian calculations with both the widely-used relativistic Cowling approximation and the linearized relativistic formulation for the f-mode of 3 analytic solutions and also the SLy4 EOS. Due to the finite densities at the NS surface, the analytic incompressible (Inc) solution is manifestly different from those of the analytic Tolman VII (T VII) and Buchdahl (Buch) as well as the SLy4 EOS. Although both the Newtonian and approximate Cowling calculations tend to overestimate f-mode frequencies, the Newtonian calculation performs extremely well in low mass NSs and better than the Cowling approximation in canonical mass NSs.  However, the Newtonian approximation becomes worse than the Cowling approximation in the highest mass NSs. This is reasonable since the Newtonian calculation keeps gravitational perturbations, which the Cowling approximation ignores. However, for massive and extremely compact NSs, relativistic corrections overwhelm the corrections due to gravitational perturbations.
 Since we require accurate results to formulate EOS-insensitive and quasi-universal relations with the compactness, tidal deformability and/or moment of inertia, further calculations are only performed in the full linearized general relativistic limit.
 
We next calculated the f-mode frequencies and damping times of hadronic, hybrid and pure quark (self-bound) NSs employing  parameterized EOSs.  Always enforcing both causality and lower and upper maximum mass limits (i.e., $2M_\odot<M_{\rm max}<2.6M_\odot$), f-mode frequencies lie in the range 1.3-2.8 kHz and damping times in the range 0.1-1 s for all configurations. The f-mode frequency of pure quark stars with canonical masses depends relatively weakly on mass, similarly to the mass-dependence of the radius of hadronic NSs. Whereas the f-mode frequencies of hadronic NSs increase smoothly with mass, hybrid NSs with strong first order transitions can result in twin star configurations which have different f-mode frequencies for the same mass. We note that low mass stars ($\approx$ 1 M$_\odot$) with high f-mode frequencies can only be achieved without the existence of a crust, i.e., only for self-bound (pure quark) stars, just as the radius of a $1M_\odot$ NS with a normal crust cannot be smaller than $\approx 10.5$ km. If such large frequencies are observed, it would be an indication of a very small radius, and evidence for the existence of pure quark stars.

Employing a range of parameterizations covering the allowed physical limits, we find that the dimensionless f-mode frequency is proportional to $\beta^{5/4}$ for hadronic and hybrid NS. In contrast, pure quark stars follow $\Omega_f\propto \beta^{3/2}$ which is found for analytic solutions in the Newtonian approximation. We also verify that the f-mode correlates strongly (to better than 1\%) with other radial moments of NSs, even for hybrid and pure quark stars. These are known as the $\bar I-\Lambda-\Omega_f$ relations \cite{lau2010inferring}.  We note that the $\Lambda$-$\Omega_f$ correlation is slightly more accurate than the $\bar I-\Omega_f$ correlation, reaching 0.1\% accuracy expect for low mass hybrid NSs with large quark cores. We provide fitting formula for these universal relations, as well as less accurate EOS-insensitive correlations with the compactness $\beta$. Our fitting formulae agree with previous works but are more accurate and applicable to wider range of NS masses and EOSs.

We discovered an abnormal f-mode in hybrid NSs displaying the twin star phenomenon with central pressures just above the quark-hadron transition pressure $p_t$.  These exhibit  manifestly different profiles of fluid and metric perturbation amplitudes. Canonical f-modes have no nodes for both fluid and metric perturbation amplitudes, e.g. Fig. \ref{fig:HKWX_EOS_SLY4_1.4}, left upper panel, and Fig. \ref{fig:HKWX_hybrid_branches}, lower two panels. As the quark core first develops, the amplitude $V$ becomes discontinuous and flips its sign at the quark-hadron interface, while the amplitudes $H_0$, $X$ and $W$ form nodes close to the interface (the 1-node II state). With increasing central pressure and quark core size, the nodes in $H_0$, $X$ and $W$ eventually disappear, while $V$ simultaneously forms a node (the 1-node I state. Further increases in central pressure lead to the disappearance of the node in $V$ and normal 0-node behavior is restored.  In both 1-node cases, the f-mode frequency moderately deviates from the $\bar I-\Lambda-\Omega_f$ universal relation (the only violations we have found), and also a gap forms in the f-mode frequency spectrum.  This might provide an opportunity to directly observe the existence of a strong first order transition.  Although the example we show in Fig. \ref{fig:HKWX_hybrid_branches} is outside of the canonical NS mass range, having $M_t<1M_\odot$ which would prove to be unobservable in practice, the use of a stiffer EOS above about $1.5-2n_s$ (even while continuing to satisfy the tidal deformability constraint from GW170817, neutron matter and nuclear systematic constraints near $n_s$, and the $M_{\rm max}>2M_\odot$ constraint) could provide suitable conditions for obtaining the twin star phenomenon with $M_t>1M_\odot$. To our knowledge, we are the first to report the existence of this special 1-node behavior of the f-mode.

The discontinuous g-mode frequency (which requires the existence of a discontinuity in the density) depends strongly on the magnitudes of both the transition density and its discontinuity.  On the other hand, for a given set of EOS parameters, it depends weakly on the stellar mass. Uncertainties in the low-density hadronic EOS contribute less than about 5\% uncertainty to the g-mode frequency. Due to causality and maximum mass constraints, the discontinuous g-mode frequency has an upper bound of about 1.5 kHz. However, if the squared sound speed in the inner core is restricted to $c^2\le1/3$, the discontinuous g-mode can only reach about 0.8 kHz, which is significantly lower than the f-mode frequency of 1.3-2.8 kHz. Also, in this eventuality, the g-mode gravitational wave damping time is usually extremely long, being larger than $10^4$ s compared to about $10^2$ s for an inner core in which $c_s^2\le1$, and is also large compared with the usual f-mode damping time, 0.1-1 s. 
We found an improved fit for the g-mode frequency that depends only on $M/R,M_t/M,R_t/R$ and $\Delta\varepsilon/\varepsilon_t$ and which is otherwise insensitive to the hybrid EOS parameters $\varepsilon_t$ and $c_s^2$ as well as the assumed hadronic EOS.
  We found the Cowling approximation is accurate to within 20\%, which is significantly less accurate than previously estimated, due to the consideration of a larger variety of first order transitions and realistic NS masses.  Our analytic approximation has a similar accuracy.

In this work, we assume the perturbed fluid is ideal. Superfluidity inside the NS introduces an additional flow component which is discussed in other works, such as Ref.  \cite{Gualtieri2014}. The fluid perturbations are also assumed to be barotropic, which holds only when matter is adiabatic and always in equilibrium except for the phase conversion between quarks and hadrons. Non-barotropic EOSs involving composition or temperature gradients could lead to modification of our present results and this warrants further investigation. Non-adiabatic effects, such as from the neutrino Urca processes during the inspiral and merger phases, could introduce significant additional damping due to bulk viscosity~\cite{alford2020bulk,arras2019urca}.

\begin{acknowledgments}

T.Z is supported by the U.S. DOE Grant No. DE-FG02-93ER40756. J.M.L. and T.Z. acknowledge support by the U.S. DOE under Grant No. DE-FG02-87ER40317 and by NASA's NICER mission with Grant 80NSSC17K0554. 

\end{acknowledgments}

\appendix
\section{Piecewise Polytrope Crust EOS \label{sec:crustEOS}}

\begin{table}[htbp!]
\caption{Piecewise polytropic EOS parameter set fitting the SLy4 crust EOS} 
\begin{center} 
\begin{tabular}{|c|cccc|}
\hline
i & $~n_i$ (fm$^{-3}$)  & $~\varepsilon_i$ (MeV fm$^{-3}$) &  $~p_i$ (MeV fm$^{-3}$)   & $~\gamma_{i-1}$\\ \hline
~0~  &  2$\times 10^{-6}$  &  1.8658 $\times 10^{-3}$  &  1.1763 $\times 10^{-6}$  &  1.35 \\
1  &  0.0003  &  0.28163  &  6.1751 $\times 10^{-4}$ &  5/4 \\
2  &  0.0013  &  1.2232  &  1.2855 $\times 10^{-3}$ &  1/2 \\
3  &  0.04  &  37.889  &  0.12394  &  4/3 \\
\hline
\end{tabular}
\end{center}
\label{tab:piecewise_polytrop_crust_parameter}
\end{table}

The requirements of high accuracy (due to the small imaginary part of the quasi-normal oscillation frequency) and reasonable computation times suggests the use of an analytical crust EOS to reduce interpolation error.  The crust EOS used in this study is a piecewise polytrope EOS with 4 segments, with adiabatic indices 1.35, 5/4, 1/2, and 4/3 chosen to mimic the Sly4 crustal EOS. The 1st and 2nd segments correspond to the NS outer crust. The 3rd segment represents the region where the electron fraction decreases rapidly, from about 0.4 to 0.04, with increasing density. In the 4th segment, the electron fraction remains nearly constant, and neutron pressure dominates electron pressure. This segment terminates at the core-crust boundary, fixed at $n_B=0.04$ fm$^{-3}$, smoothly matching the Sly4 energy density and pressure at that density, namely $\varepsilon=37.88$ MeV fm$^{-3}$ and $p=0.1239$ MeV fm$^{-3}$, respectively.  Densities, pressures and polytropic indices at the boundaries of the crustal piecewise poplytrope segments are presented in Table \ref{tab:piecewise_polytrop_crust_parameter}.

\begin{figure}
    \centering
    \includegraphics[width=\linewidth]{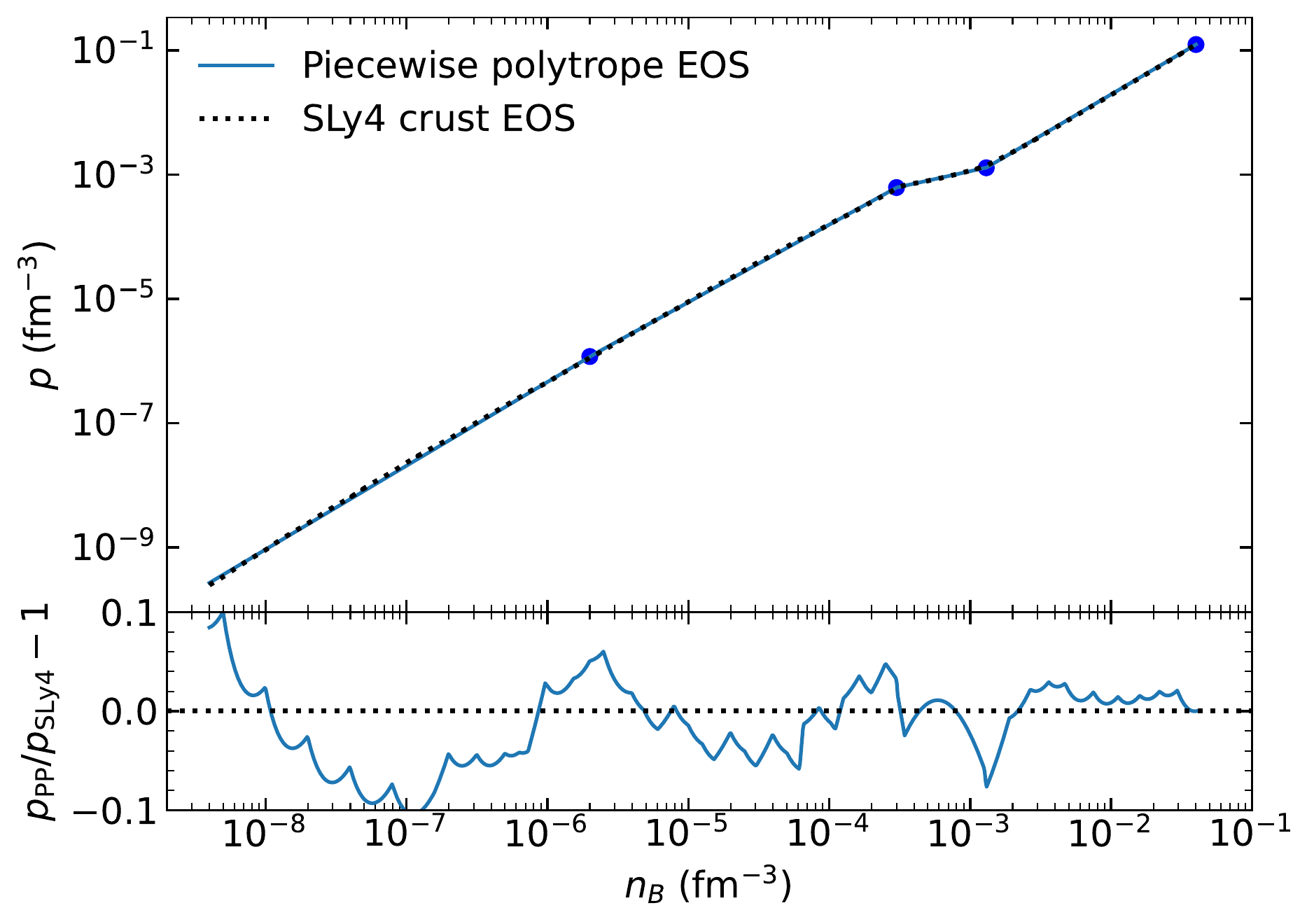}
    \caption{Comparing the energy density as a function of baryon density for piecewise polytrope and SLy4 NS crust EOSs. Blue dots correspond to boundaries of the piecewise polytrope EOS in Table \ref{tab:piecewise_polytrop_crust_parameter}.\label{fig:crust_eos_compare}}
\end{figure}

Fig. \ref{fig:crust_eos_compare} compares the crust piecewise polytrope and SLy4 EOSs. The relative error of the piecewise-polytropic approximation is generally less than 10\% for $n_B>10^{-10}$ fm$^{-3}$. The piecewise polytrope more than adequately portrays the EOS from the core-crust boundary at $n_B=0.04$ fm$^{-3}$ to vacuum at the surface, where the energy per baryon is 931.2 MeV. We verified that the masses and radii, as functions of central density or pressure, employing this piecewise-polytropic crust EOS agree with those employing the full SLy4 EOS to within 0.03\% and 0.001\%, respectively, for canonical mass NSs. 


\nocite{*}

\bibliography{ref}

\end{document}